\documentclass[aps,prb,amsmath,amssymb,nofootinbib,10pt,notitlepage,twocolumn]{revtex4-2}

\usepackage[T2A]{fontenc}
\usepackage[utf8]{inputenc}

\usepackage{graphicx}
\usepackage[dvipsnames]{xcolor}
\usepackage[colorlinks=true, allcolors=blue]{hyperref}
\usepackage{braket}
\usepackage{suffix}
\usepackage{xparse}
\usepackage{mathtools}
\mathtoolsset{multlined-width=0.95\displaywidth}

\newcommand{\dd}{\mathrm{d}}
\newcommand{\thin}{\mkern1.5mu}

\newcommand{\V}[1]{\boldsymbol{#1}}  
\newcommand{\M}[1]{\hat{#1}}  

\newcommand{\FM}[1]{\widecheck{\mathcal{#1}}}  

\usepackage{latexml}
\iflatexml
    \newcommand{\widecheck}[1]{\check{#1}}
\else
    \DeclareFontFamily{U}{mathx}{}
    \DeclareFontShape{U}{mathx}{m}{n}{<-> mathx10}{}
    \DeclareSymbolFont{mathx}{U}{mathx}{m}{n}
    \DeclareMathAccent{\widehat}{0}{mathx}{"70}
    \DeclareMathAccent{\widecheck}{0}{mathx}{"71}
\fi

\def\dbar{{\mkern3mu\mathchar'26\mkern-12mu d}}

\NewDocumentEnvironment{subalign}{ob}{
\begin{subequations}
    \IfValueT{#1}{\label{#1}}
    \begin{align}
        #2
    \end{align}
\end{subequations}}
{\ignorespacesafterend}

\NewDocumentEnvironment{spgather}{b}{
\begin{equation}
    \begin{gathered}
        #1
    \end{gathered}%
\end{equation}}
{\ignorespacesafterend}

\NewDocumentEnvironment{spalign}{b}{
\begin{align}
    \begin{split}
        #1
    \end{split}%
\end{align}}
{\ignorespacesafterend}

\NewDocumentEnvironment{spalignat}{mb}{
\begin{equation}
    \begin{alignedat}{#1}
        #2
    \end{alignedat}
\end{equation}
}
{\ignorespacesafterend}

\renewcommand{\aa}{\mathsf{a}}
\newcommand{\bb}{\mathsf{b}}

\newcommand{\dof}{\mathsf{dof}}

\newcommand{\KK}{\mathsf{K}}
\newcommand{\GG}{\mathsf{G}}

\newcommand{\MX}{\M{X}}
\newcommand{\MXb}{\MX_\bb}
\newcommand{\MY}{\M{Y}}
\newcommand{\MYb}{\MY_\bb}

\newcommand{\MA}{\M{A}}
\newcommand{\MB}{\M{B}}
\newcommand{\MW}{\M{W}}
\newcommand{\MV}{\M{V}}

\newcommand{\MG}{\M{G}}
\NewDocumentCommand{\MGb}{e{_^}}{\M{G}_{\bb\IfValueT{#1}{#1}}\IfValueT{#2}{^{#2}}}

\newcommand{\Mg}{\M{g}}
\NewDocumentCommand{\Mgb}{e{_^}}{\M{g}_{\bb\IfValueT{#1}{#1}}\IfValueT{#2}{^{#2}}}

\NewDocumentCommand{\MSIb}{e{_^}}{\M{Σ}_{\bb\IfValueT{#1}{#1}}\IfValueT{#2}{^{#2}}}

\newcommand{\MS}{\M{S}}
\NewDocumentCommand{\MSb}{e{^_}}{\M{S}^{\bb\IfValueT{#1}{#1}}\IfValueT{#2}{_{#2}}}

\newcommand{\MT}{\M{T}}
\NewDocumentCommand{\MTb}{e{^_}}{\M{T}^{\bb\IfValueT{#1}{#1}}\IfValueT{#2}{_{#2}}}

\newcommand{\FMT}{\FM{T}}
\NewDocumentCommand{\FMTb}{e{^_}}{\FM{T}^{\bb\IfValueT{#1}{#1}}\IfValueT{#2}{_{#2}}}

\newcommand{\FMK}{\FM K}
\NewDocumentCommand{\FMKb}{e{_^}}{\FM{K}_{\bb\IfValueT{#1}{#1}}\IfValueT{#2}{^{#2}}}

\newcommand{\FMC}{\FM C}

\newcommand{\FMR}{\FM R}
\NewDocumentCommand{\FMRb}{e{_^}}{\FM{R}_{\bb\IfValueT{#1}{#1}}\IfValueT{#2}{^{#2}}}

\DeclareMathSymbol{:}{\mathbin}{operators}{"3A} 

\newcommand{\Tr}[1]{\operatorname{Tr}\bigl[#1\bigr]}
\newcommand{\Trr}[1]{\operatorname{Tr}\biggl[#1\biggr]}

\usepackage{textgreek}

\newcommand{\Vr}{{\V{r}}}

\newcommand{\Vp}{\V{p}}

\newcommand{\eff}{\mathsf{eff}}

\usepackage{mathrsfs} 
\usepackage{common-unicode} 

\newcommand{\avg}[1]{\left<#1\right>}
\newcommand{\avgs}[1]{\bigl<#1\bigr>}

\newcommand{\newdiag}[2][-2pt]{\mathord{\raisebox{#1}{\includegraphics[page=#2,scale=0.25]{new_diags-5.pdf}}}}

\begin{document}

\newcommand{\YaroslavNote}[1]{{\textcolor{blue}{{\bf Yaroslav:} \it #1}}}
\newcommand{\Yaroslav}[1]{{\color{blue} #1}}

\newcommand{\ConyuhNote}[1]{{\textcolor{violet}{{\bf Conyuh:} \it #1}}}
\newcommand{\Conyuh}[1]{{\color{violet} #1}}

\newcommand{\IgorNote}[1]{{\textcolor{orange}{{\bf Igor:} \it #1}}}
\newcommand{\Igor}[1]{{\color{orange} #1}}

\title{Nanoscale linear response of strongly disordered stable solids}
\date{\today}
\author{D.\,V.~Babin}
\author{I.\,O.~Raikov}
\author{Y.\,M.~Beltukov}
\email{yaroslav.beltukov@mail.ioffe.ru}
\affiliation{Ioffe Institute, Politechnicheskaya st.~26, 194021 St.~Petersburg, Russia}

\begin{abstract}
    Strongly disordered solids exhibit distinctive properties at the nanoscale, where conventional continuum theory breaks down. We show that their disorder-averaged linear response can be described by a modified continuum theory in which the response coefficients are local but depend nonlocally on the structural properties. The stability criterion requires the response operator to be positive semidefinite, which naturally leads to a correlated Wishart disorder. In the limit of strong disorder, the resulting theory reduces the long-wavelength response to a scalar disorder-induced contrast field. In elasticity, this field describes the formation of a stiffened    interphase around rigid nanoparticles and boundaries, whose characteristic extent is set by the nonaffine length. The predictions are confirmed by molecular-dynamics simulations of a Lennard-Jones glass and a model polymer. Direct calculations of the nonlocal elastic kernels show that their spatial range is much shorter than the extent of the stiffened interphase, demonstrating that the latter does not require long-range constitutive nonlocality.
    
\end{abstract}

\maketitle

\section{Introduction}
\label{sec:Intro}

Amorphous and structurally disordered solids constitute a broad class of condensed-matter systems, including inorganic and metallic glasses, glassy polymers, amorphous semiconductors, disordered composite materials, as well as porous and granular media. Their mechanical, dielectric, thermal, and transport properties are important both for fundamental condensed-matter physics and for applications ranging from structural materials and coatings to electronic, photonic, and energy technologies~\cite{
    Elliott-physics-amorphous-materials-1984,
    Torquato-random-heterogeneous-materials-2002,
    Binder-glassy-materials-disordered-2005,
    Alexander-amorphous-solids-their-1998,
    Zaccone-theory-disordered-solids-2023}.

Despite their chemical and structural diversity, disordered solids share the absence of long-range order. Consequently, their microscopic environments are spatially nonuniform, and the connection between microscopic structure and macroscopic properties is considerably less direct than in crystalline materials. Structural disorder affects different linear-response phenomena in distinct but closely related ways. Under mechanical loading, a macroscopically homogeneous deformation field gives rise to irregular microscopic particle displacements that deviate from the corresponding affine deformation, commonly referred to as \emph{nonaffine} displacements \cite{
    Lemaitre-sum-rules-quasistatic-2006,
    Zaccone-approximate-analytical-description-2011,
    Szamel-elastic-constants-zerotemperature-2023}.
Analogously, in the context of dielectric response, a spatially uniform external electric field induces spatially heterogeneous electronic polarization and ionic displacements, which are determined by the local atomic environments and their coupling to the applied field \cite{
    Zhao-structural-electronic-dielectric-2005,
    Ceresoli-structural-dielectric-properties-2006,
    Momida-theoretical-study-dielectric-2006}.
Disorder also profoundly modifies electronic, thermal, and wave transport properties \cite{
    Mott-electronic-processes-noncrystalline-2012,
    Allen-thermal-conductivity-disordered-1993,
    Sheng-introduction-wave-scattering-2006,
    Kirkpatrick-classical-transport-disordered-1971,
    Zhou-thermal-conductivity-amorphous-2020}.
In all these cases, the macroscopic response emerges from spatially heterogeneous and collectively coupled microscopic degrees of freedom. A general description of disordered media must therefore account for the heterogeneous, collective, and potentially nonlocal nature of their linear response.

At sufficiently large length scales, microscopic fluctuations are averaged out, and a disordered material can often be characterized by homogeneous macroscopic response coefficients. At the nanoscale, however, the characteristic dimensions of the observation volume, probe, inclusion, or interfacial region may become comparable to the characteristic scale of structural and response heterogeneity. Under these conditions, local elastic moduli, dielectric susceptibilities, and transport coefficients may vary substantially from one region to another. In addition, correlations between spatially separated regions may give rise to genuinely nonlocal constitutive response \cite{
    Baumgarten-nonlocal-elasticity-jamming-2017,
    Reda-gradient-mechanical-properties-2022}.
A spatially resolved theory of disorder-averaged linear response is therefore needed to determine how macroscopic constitutive behavior emerges from heterogeneous microscopic degrees of freedom and to describe systems for which conventional continuum or effective-medium approximations are no longer sufficient
\cite{
    Tanguy-continuum-limit-amorphous-2002,
    Yoshimoto-mechanical-heterogeneities-model-2004,
    Wagner-local-elastic-properties-2011,
    Mizuno-measuring-spatial-distribution-2013,
    Rodney-modeling-mechanics-amorphous-2011,
    Choy-effective-medium-theory-2015}.

The microscopic elastic response of amorphous solids remains heterogeneous over length scales that can substantially exceed the interatomic distance. Previous numerical and theoretical studies have shown that continuum elasticity is recovered only above a material-dependent heterogeneity scale, whereas below this scale the local elastic moduli and displacement fields exhibit pronounced spatial fluctuations
\cite{
    Tanguy-continuum-limit-amorphous-2002,
    Leonforte-continuum-limit-amorphous-2005,
    Goldenberg-particle-displacements-elastic-2007,
    DiDonna-nonaffine-correlations-random-2005,
    Tanguy-vibration-modes-characteristic-2015,
    Lerner-breakdown-continuum-elasticity-2014,
    Lerner-anomalous-linear-elasticity-2023}.
Recent work demonstrated that the nonaffine displacement field possesses large-scale exponential correlations characterized by a nonaffine length $ξ$, which can reach many interatomic distances and substantially exceed the structural correlation length \cite{
    Conyuh-largescale-exponential-correlations-2026}.
Thus, $ξ$ characterizes the spatial extent of correlated local strain and rotation fluctuations and separates the microscopic heterogeneous regime from the scales at which classical continuum elasticity becomes applicable. In what follows, we refer to the regime in which $\xi$ substantially exceeds the microscopic structural scale as the strongly disordered regime.

Relaxation of particle positions away from the affine deformation makes an essential contribution to the elastic response of amorphous solids and generally reduces their stiffness relative to the affine approximation
\cite{
    Lemaitre-sum-rules-quasistatic-2006,
    Vaibhav-timescale-bridging-atomistic-2024,
    Zaccone-approximate-analytical-description-2011}.
Consequently, constraints imposed by a rigid boundary or inclusion can suppress nonaffine motion and produce an interfacial region with enhanced elastic moduli. Molecular-dynamics simulations of polystyrene containing silica nanoparticles revealed such an elastically enhanced region extending approximately 1.4 nm from the nanoparticle, considerably farther than the associated perturbation of the polymer density \cite{
    Beltukov-local-elastic-properties-2022}.
Related simulations and experiments have also revealed graded
interfacial elastic properties and nanoscale mechanical reinforcement
in polymer nanocomposites
\cite{
    Cheng-unraveling-mechanism-nanoscale-2016,
    Brune-direct-measurement-rubber-2016,
    Fankhanel-elastic-interphase-properties-2019,
    Barakat-predicting-mechanical-heterogeneity-2024}.
Theoretical analysis showed that an interphase of thickness of order $ξ$ can arise solely from the suppression of the nonaffine deformation field, without requiring a comparably extended structural perturbation \cite{
    Conyuh-effective-elastic-moduli-2023}.
Because the relative volume of this interphase increases as the inclusion size decreases, the effect becomes particularly important when the nanoparticle radius is comparable to $ξ$. It may then substantially enhance the influence of the inclusions on the macroscopic stiffness and provide a microscopic basis for the interphase introduced phenomenologically in three-phase models \cite{
    Odegard-modeling-mechanical-properties-2005,
    Qiao-simulation-interphase-percolation-2009,
    Huang-interphase-polymer-nanocomposites-2022}.
An important question is whether the spatial extent of such an interphase reflects an intrinsically nonlocal elastic response of comparable range, or whether the constitutive response remains local while its effective coefficients vary over the larger collective scale $ξ$.

In the present work, the spatially resolved response is considered in a disorder-averaged sense \cite{
    Choy-effective-medium-theory-2015,
    Feng-effectivemedium-theory-percolation-1985,
    Zhou-general-theory-calculating-2017}.
For a particular realization of an amorphous solid, the local response exhibits strong microscopic fluctuations associated with its specific atomic configuration. We consider instead an ensemble of systems with the same statistical properties, macroscopic geometry, and boundary conditions, but different realizations of structural disorder. In a homogeneous bulk medium, this averaging restores translational invariance and produces spatially uniform effective moduli. In the presence of an interface or inclusion, however, translational invariance is explicitly broken, and the disorder-averaged response remains position dependent. The effective contrast $α(\Vr)$ introduced below describes this systematic spatial variation rather than the microscopic fluctuations of an individual realization.

Importantly, disorder averaging does not eliminate the effect of nonaffine displacements \cite{
    Lemaitre-sum-rules-quasistatic-2006,
    Zaccone-approximate-analytical-description-2011,
    Sussman-strain-fluctuations-elastic-2015}.
Although the nonaffine displacement field may have zero ensemble average in a statistically homogeneous system, the relaxation of particle positions away from the affine deformation lowers the elastic energy and thereby reduces the disorder-averaged elastic moduli relative to the affine approximation. When these nonaffine displacements are constrained near a rigid boundary or inclusion, this softening is partially suppressed. The resulting enhancement of the effective modulus survives ensemble averaging and gives rise to a smooth spatial profile of the elastic contrast.

Random-matrix approaches provide a statistical framework for describing the vibrational and mechanical properties of amorphous solids and have been used, in particular, to describe their vibrational spectra and the boson peak~\cite{
    Grigera-vibrations-glasses-euclidean-2002,
    Manning-random-matrix-definition-2015,
    Beltukov-iofferegel-criterion-diffusion-2013,
    Baggioli-vibrational-density-states-2019,
    Conyuh-random-matrix-approach-2021}.
For a mechanically stable system, however, the static response operator must be positive semidefinite, which imposes an important constraint on an admissible random-matrix ensemble. A natural way to satisfy this condition is to represent the disordered contribution in a Wishart form~\cite{
    Beltukov-iofferegel-criterion-diffusion-2013,
    Conyuh-random-matrix-approach-2021}.

A spatial theory of the interphase was previously obtained within a scalar random-matrix model in which different bonds were assumed to be statistically independent \cite{
    Conyuh-effective-elastic-moduli-2023}.
Let $α(\Vr)$ denote the dimensionless excess elastic contrast, defined so that $α=0$ in the homogeneous bulk. Its spatial distribution satisfies \cite{
    Conyuh-effective-elastic-moduli-2023}
\begin{equation}
    α(\Vr)
    =
    ξ^2 Δ \ln\!\left[1+α(\Vr)\right],
    \label{eq:alpha1}
\end{equation}
where $Δ$ is the Laplacian. The equation contains $ξ$ as its only intrinsic length parameter, while the geometry and mechanical coupling to the surrounding bodies are specified through the boundary conditions. Although Eq.~(\ref{eq:alpha1}) is nonlinear in $α$, this nonlinearity does not represent nonlinear elasticity: it results from the self-consistent averaging of a disordered linear-response operator.

The derivation of Eq.~(\ref{eq:alpha1}) in Ref.~\cite{
    Conyuh-effective-elastic-moduli-2023}
relied on the scalar approximation and the assumption of statistically independent bonds. The aim of this paper is to establish to what extent this simple equation remains applicable under substantially more general conditions. We show that the long-wavelength response of a broad class of stable strongly disordered systems is still governed by a single scalar contrast field. For finite contrasts, the general theory yields a modified nonlinear equation containing one additional parameter $\eta$, with Eq.~(\ref{eq:alpha1}) recovered exactly for $\eta=1$. Except for limiting boundary conditions, the dependence of the spatial profiles on $\eta$ is weak, so that Eq.~(\ref{eq:alpha1}) remains a good approximation in a broad range of physically relevant situations. We focus primarily on the static elastic response, but the underlying formulation is expressed in terms of a general positive-semidefinite disordered response operator and is not intrinsically restricted to elasticity.

The remainder of the paper is organized as follows.
Section~\ref{sec:qualitative} provides a qualitative picture of nonaffine deformations and their suppression near rigid boundaries.
In Sec.~\ref{sec:stable}, we formulate stable disordered linear-response operators in terms of a correlated Wishart ensemble.
Section~\ref{sec:green} introduces the correlation superoperator and develops the corresponding disorder-averaged Green-function formalism and Dyson equations.
In Secs.~\ref{sec:small} and~\ref{sec:finite}, we analyze weak and finite spatial perturbations, respectively, and show how the long-wavelength response reduces to a scalar disorder-induced contrast field governed by a screened Poisson equation in the weak-contrast limit and by its nonlinear generalization for finite contrast.
Section~\ref{sec:MD} tests the theory using molecular-dynamics simulations of a semiflexible model polymer and a binary Lennard-Jones glass, including a direct calculation of the nonlocal elastic kernels.
Section~\ref{sec:discussion} discusses the physical interpretation of the results, their relation to nonlocal and strain-gradient elasticity, and possible extensions beyond elasticity.
The main conclusions are summarized in Sec.~\ref{sec:conclusion}, while technical derivations are given in the appendices.

\section{Qualitative picture}
\label{sec:qualitative}

\begin{figure}
    \centering
    \includegraphics[scale=0.8]{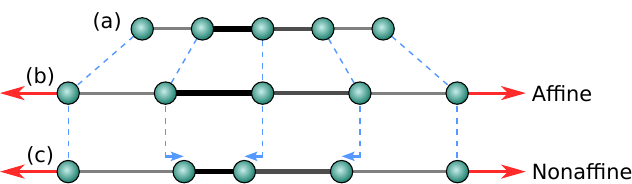}
    \caption{One-dimensional illustration of nonaffine deformation in a chain of particles connected by bonds of unequal stiffness, represented by the thickness and color of the horizontal lines. (a) Undeformed chain. (b) Affine deformation, for which all bonds undergo the same relative extension. (c) Energy-minimizing nonaffine deformation at the same total elongation of the chain. The nonaffine displacements relative to the affine configuration are indicated by blue arrows, and the dashed blue lines are visual guides.}
    \label{fig:nonaffine-1d}
\end{figure}

Before presenting the formal theory, we outline the physical mechanism underlying the spatial variation of the disorder-averaged effective response.

Figure~\ref{fig:nonaffine-1d} illustrates this mechanism for a single realization of a one-dimensional chain whose particles are connected by bonds of unequal stiffness. Under an affine deformation, all bonds undergo the same relative extension, as shown in Fig.~\ref{fig:nonaffine-1d}(b). This configuration is generally not the minimum-energy state. When the particles are allowed to relax while the total elongation of the chain is kept fixed, the deformation is redistributed: softer bonds extend more, whereas stiffer bonds extend less [Fig.~\ref{fig:nonaffine-1d}(c)]. The resulting nonaffine configuration has a lower elastic energy than the affine one.

The precise nonaffine displacement pattern depends on the particular realization of disorder. Consequently, averaging over different realizations may yield a vanishing mean nonaffine displacement. The associated reduction of elastic energy, however, does not vanish under averaging. Allowing nonaffine displacements therefore lowers the
disorder-averaged elastic modulus, whereas constraining
these displacements increases the effective stiffness.

\begin{figure}
    \centering
    \includegraphics[scale=0.8]{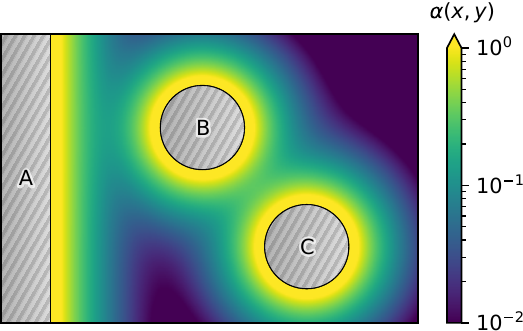}
    \caption{Two-dimensional illustration of the spatial stiffening of an amorphous matrix caused by the suppression of nonaffine deformations near rigid boundaries and inclusions, as described by Eq.~(\ref{eq:alpha1}). The hatched gray regions represent a planar rigid boundary (A) and two circular inclusions (B, C). The surrounding amorphous matrix is colored according to the effective elastic contrast $\alpha(x,y)$. The yellow regions correspond to the strongest suppression of nonaffine displacements and, consequently, to the largest stiffness enhancement. The nonaffine length is chosen as $ξ=R/2$, where $R$ is the radius of the circular inclusions.}
    \label{fig:halo}
\end{figure}

This mechanism provides a qualitative explanation for the formation of a disorder-averaged stiffened region around a rigid nanoparticle, as illustrated in Fig.~\ref{fig:halo}. The inclusion constrains the motion of nearby particles and restricts their ability to undergo nonaffine displacements. Although the microscopic displacement and stiffness fields fluctuate between different amorphous realizations, the restriction of nonaffine motion caused by the inclusion is systematic and remains after ensemble averaging. The disorder-averaged effective modulus is therefore larger near the inclusion than in the homogeneous bulk medium.

The constrained region, in turn, restricts nonaffine motion in the surrounding material. Through the mechanical coupling between neighboring regions, this constraint propagates away from the interface in a self-consistent manner. Such self-consistent propagation gives rise to the nonaffine length $\xi$, which sets the characteristic decay length of the boundary-induced constraint and, in a homogeneous bulk system, also characterizes the spatial correlations of nonaffine displacements. The resulting smooth effective contrast $\alpha(\Vr)$ gradually approaches zero over a distance of order $\xi$, as described by Eq.~(\ref{eq:alpha1}).

A similar qualitative mechanism may arise in the static dielectric response. A rigid boundary or inclusion can modify the local ionic and electronic polarization by constraining atomic displacements, changing the local interaction network, or perturbing the redistribution of electronic charge. Although the microscopic degrees of freedom differ from those governing elasticity, the effective response may again be altered over a finite region because the local polarization is collectively coupled to its surroundings.

This qualitative picture nevertheless leaves two important theoretical questions open. First, atomic displacements are vector quantities and elastic response is generally described by a fourth-order stiffness tensor. It is therefore not evident whether the spatial variation of the tensorial elastic response can still be governed by the single scalar contrast $\alpha(\Vr)$ in Eq.~(\ref{eq:alpha1}), or whether a
tensor-valued contrast field is required. The role of the vector character of atomic displacements in glassy systems has been discussed in Refs.~\cite{
    Phillips-constraint-theory-vector-1985,
    Skipetrov-anderson-transition-elastic-2018,
    Szamel-sound-attenuation-glasses-2025}.

Second, the logarithmic term makes Eq.~(\ref{eq:alpha1}) nonlinear in the effective contrast. In the limit of a weak contrast, $|α|\ll 1$, the expansion of the logarithm gives
\begin{equation}
    α(\Vr)=ξ^2 Δα(\Vr),
    \label{eq:alpha2}
\end{equation}
which is the homogeneous screened Poisson equation. For larger contrasts, however, it is not immediately clear whether the logarithmic dependence remains valid when the scalar approximation and the assumption of statistically independent bonds employed in Ref.~\cite{Conyuh-effective-elastic-moduli-2023} are relaxed, or whether additional nonlinear terms appear.

In the following sections, we show that the scalar structure of
Eq.~(\ref{eq:alpha1}), including its logarithmic nonlinearity, persists for
a broad class of strongly disordered stable systems. In particular,
no tensor-valued contrast field is required. The more general theory
retains the logarithmic term and introduces only a specific additional
term proportional to $\alpha(\Vr)/[1+\alpha(\Vr)]$, while
preserving $\xi$ as the characteristic length scale governing the
spatial variation of the effective response.

\section{Stable amorphous solids}
\label{sec:stable}

\subsection{Linear response}

Amorphous solids lack long-range order but can remain near mechanically stable metastable configurations over long time scales. In the present work, we consider an amorphous solid at a temperature $T$ well below the glass transition temperature $T_g$, where structural relaxation is slow on the time scale of the applied perturbation~\cite{
    McKenna-50th-anniversary-perspective-2017}.
The linear response can then be considered about a fixed mechanically stable configuration.

Although we formulate the theory in terms of the elastic response of
an atomic system, the same construction can be applied more generally
to stable linear-response problems with additional or different
microscopic degrees of freedom. For example, molecules or granules may
possess rotational or internal degrees of freedom, while charge and
magnetic-moment dynamics can be considered in dielectric and magnetic
response, respectively. The essential requirement is that the response
is considered about a stable configuration.

In terms of elastic response, external forces $\V f_i$ applied at a frequency $ω$ induce particle displacements $\V u_i$ from their equilibrium positions. We consider sufficiently small external forces so that the system remains within the same mechanically stable configuration and does not undergo transitions between distinct local minima. The role of rare low-energy excitations, such as two-level systems and quasilocalized vibrations~\cite{
    Phillips-amorphous-solids-lowtemperature-1981,
    Ramos-lowtemperature-thermal-vibrational-2023},
may require special consideration, and their role within the present framework is discussed separately in Ref.~\cite{
    Conyuh-spatial-structure-quasilocal}.

In the linear approximation, the elastic response $\V u_i$ is defined by the following system of linear equations:
\begin{equation}
    \sum_{jβ} \left[Φ_{iα,jβ} - ω^2m_{iα,jβ} \right]u_{jβ} = f_{iα},
    \label{eq:lin-resp}
\end{equation}
where $\M{Φ}$ is the force-constant matrix and $\M m$ is the mass matrix. In amorphous solids, each realization has different atomic positions. To perform the averaging over different realizations, we will work on the basis of the \emph{reference lattice}, which is a simple cubic lattice (or a square lattice in two dimensions) with the same number of nodes as the number of atoms $N$ in the system. The atoms are assigned one-to-one to the reference-lattice nodes using the mapping described in Ref.~\cite{
    Conyuh-largescale-exponential-correlations-2026}.
The mapping is chosen to preserve spatial locality, so that atoms that are close in the amorphous configuration are associated with nearby reference nodes.

Therefore, the indices $i$ and $j$ in Eq.~(\ref{eq:lin-resp}) enumerate the reference nodes ($1\ldots N$), while $α$ and $β$ denote the degrees of freedom of each node. For the atomic system under consideration, they are the Cartesian indices of translational motion ($x,y,z$ for $\dbar=3$ or $x,y$ for $\dbar=2$). The elements of the force-constant matrix $\M{Φ}$ are defined as the Hessian of the total potential energy $U$ with respect to atomic displacements, expressed in terms of the displacements $u_{iα}$ at the reference nodes:
\begin{equation}
    Φ_{iα,jβ} = \frac{∂^2 U}{∂u_{iα}∂u_{jβ}^*}.
    \label{eq:Phi}
\end{equation}
Here the derivatives with respect to the generally complex variables $u_{iα}$ and $u_{jβ}^*$ are understood in the Wirtinger sense. With this convention, the harmonic energy is $U = \sum_{iαjβ} Φ_{iα,jβ} u_{iα} u_{jβ}^*$. Although static physical displacements can be chosen real, complex displacement amplitudes arise naturally in the frequency-domain formulation. We therefore retain the more general complex-valued notation, which also facilitates extensions to other linear-response problems with intrinsically complex variables.

The mass matrix $\M m$ is the Hessian of the kinetic energy with respect to the corresponding velocities $\dot{u}_{iα}$. Utilizing Eq.~(\ref{eq:lin-resp}), the atomic displacements can be explicitly expressed as
\begin{equation}
    u_{iα} = \sum_{jβ} \left(\frac{1}{\M{Φ} - ω^2\M m}\right)_{iα,jβ}f_{jβ}^{}.
\end{equation}

\subsection{Averaging}

The main goal of this paper is to study the disorder-averaged response of amorphous solids
\begin{equation}
    ⟨u_{iα}⟩ = \sum_{jβ} G_{iα,jβ} f_{jβ}^{}
\end{equation}
given by the ensemble-averaged Green function
\begin{equation}
    \MG = \avg{\big(\M{Φ} - \M m ω^2\big)^{-1}},
    \label{eq:G}
\end{equation}
where the angle brackets denote ensemble averaging. The correlation properties of nonaffine displacements given by $⟨u_{iα}u_{jβ}^*⟩$ are studied in detail in Ref.~\cite{Conyuh-largescale-exponential-correlations-2026}. In this paper, we study the ensemble-averaged response $⟨u_{iα}⟩$ and how the presence of interfaces or nanoinclusions may modify this response. To better understand the behavior of the ensemble-averaged response, we introduce an effective force-constant matrix
\begin{equation}
    \M{Φ}^\eff = \MG^{-1} + \M m ω^2
    \label{eq:Phi_eff_def}
\end{equation}
that satisfies
\begin{equation}
    \MG = \big(\M{Φ}^\eff - \M m ω^2\big)^{-1}
\end{equation}
for the ensemble-averaged Green function $\MG$. This relation means that the \emph{average response} of the disordered amorphous solid is equivalent to the deterministic response of an effective, nonrandom medium characterized by $\M{Φ}^\eff$~\cite{
    Conyuh-effective-elastic-moduli-2023}.
In field-theoretical terminology, $\M{Φ}^\eff$ is analogous to the self-energy~\cite{
    Rammer-quantum-transport-theory-2018,
    Sadovskii-diagrammatics-lectures-selected-2006}.

\subsection{Correlated Wishart disorder}

A central requirement in applying random matrix theory to the linear response of amorphous solids is mechanical stability. Within the harmonic approximation, stability requires the Hessian, or force-constant matrix, $\M{Φ}$ to be positive semidefinite. Consequently, the total potential energy
\begin{equation}
    U = \sum_{iαjβ} Φ_{iα,jβ} u_{iα} u_{jβ}^*
\end{equation}
is nonnegative for any displacements $u_{iα}$. In addition, the force-constant matrix is Hermitian, $Φ_{iα,jβ} = Φ_{jβ,iα}^*$, as follows directly from its definition in Eq.~(\ref{eq:Phi}).

Any Hermitian positive-semidefinite matrix $\M{Φ}$ admits a factorization of the form $\M{Φ}=\MB\MB^\dag$. Conversely, $\MB\MB^\dag$ is Hermitian positive semidefinite for any matrix $\MB$ \cite{Horn-matrix-analysis-2017}. In component form, this factorization reads
\begin{equation}
    Φ_{iα,jβ} = \sum_{k} B_{iα,k} B_{jβ,k}^*.
    \label{eq:AAT}
\end{equation}
Physically, this factorization expresses the harmonic potential energy of a stable system as a nonnegative sum of squared amplitudes:
\begin{equation}
    U = \sum_kU_k, \quad U_k = \biggl|\sum_{iα} B_{iα,k} u_{iα}\biggr|^2. \label{eq:Uk}
\end{equation}

The matrix $\MB$ in Eq.~(\ref{eq:AAT}) can be interpreted in terms of generalized bonds. The composite index $iα$ enumerates the degrees of freedom, whereas the index $k$ enumerates the contributions $U_k$ associated with individual generalized bonds. Each such bond may couple several degrees of freedom rather than only a pair of atoms. The factorization may be chosen to reflect the locality of the underlying interactions, so that each generalized bond couples a spatially localized set of degrees of freedom. Accordingly, the number and spatial arrangement of the nonzero elements of $\MB$ depend on the form of the interatomic interactions in the amorphous solid~\cite{Beltukov-boson-peak-various-2016}. 

The positive-semidefinite character of the bond contributions in Eq.~(\ref{eq:Uk}) should not be interpreted as implying that every individual pairwise interaction between atoms is itself positive semidefinite. Even in a mechanically stable medium, some interactions may have negative second derivatives and therefore provide destabilizing contributions to the Hessian $\M{Φ}$. Nevertheless, because the full Hessian is positive semidefinite, its quadratic form can always be decomposed into an alternative set of individually nonnegative contributions. Further details of this construction are given in Appendix~F of Ref.~\cite{Conyuh-largescale-exponential-correlations-2026}.

Motivated by the factorization in Eq.~\eqref{eq:AAT}, we represent
the force-constant matrix as
\begin{equation}
    \M{Φ} = \MA\MA^\dagger + \M{Φ}_0,
    \label{eq:MPhi}
\end{equation}
where $\MA$ will describe the disordered contribution and
$\M{Φ}_0$ is an arbitrary deterministic positive-semidefinite
contribution included for generality. Since both terms are positive
semidefinite, $\M{Φ}$ is positive semidefinite for every realization. For example, $\M{Φ}_0$ may be chosen to describe nonrandom longitudinal deformations, while the random matrix $\MA$ accounts for disorder in the transverse response. Such a decomposition is physically natural for amorphous solids, and the fluctuating shear modulus is used in heterogeneous elasticity theory (HET) \cite{
    Schirmacher-heterogeneous-elasticity-tale-2022}.
The matrix $\MA$ is generally rectangular, with dimensions $N_\dof \times N_\bb$, where $N_\dof = \dbar \thin N$ is the total number of degrees of freedom and $N_\bb$ is the number of random generalized bonds.

Following Ref.~\cite{Conyuh-effective-elastic-moduli-2023}, we assume the nonzero elements of the matrix $\MA$ to be zero-mean correlated Gaussian variables with covariance
\begin{equation}
    \avg{A_{iα,k} A_{jβ,l}^*} = \mathcal{C}_{iαjβ;kl}.
    \label{eq:C-corr}
\end{equation}
The random contribution $\MA\MA^\dagger$ therefore forms a correlated
Wishart ensemble \cite{
    Sengupta-distributions-singular-values-1999,
    Burda-spectral-moments-correlated-2005,
    Vinayak-correlated-wishart-ensembles-2010}.
Matrices of this form are also referred to as random Gram matrices \cite{
    Alt-local-law-random-2017}.
In the isotropic Gaussian case, where the elements of $\MA$ are independent and identically distributed, the ensemble reduces to the classical Wishart–Laguerre ensemble \cite{
    Livan-introduction-random-matrices-2018}.
There is also a closely related chiral ensemble~\cite{
    Forrester-loggases-random-matrices-2010}.

For conventional elastic systems with a real symmetric force-constant
matrix, $\MA$ may be chosen real. In the present work, however, we consider the more general complex-valued case, for which matrix transposition must be distinguished from Hermitian conjugation. A general complex Gaussian matrix $\MA$ is additionally characterized
by the covariance $\avg{A_{iα,k}A_{jβ,l}}$, which is not determined
by the covariance in Eq.~(\ref{eq:C-corr}). This second correlation
does not contribute to the effects considered here, as shown in
Appendix~\ref{app:Dyson}.

We refer to this class of disorder, defined by $\M{Φ}$ in Eq.~(\ref{eq:MPhi}) with the underlying Gaussian matrix $\MA$, as \emph{Wishart disorder}. By construction, this type of disorder ensures that $\M{Φ}$ is positive semidefinite, in contrast to the conventional Gaussian disorder, which does not in general guarantee positive semidefiniteness. In this work, we employ field-theoretical techniques to investigate the properties of Wishart disorder.

\section{Green functions}
\label{sec:green}

\subsection{Correlation superoperator}

Before we proceed, we need to introduce the main notation used in this paper. We will use the language of operators and superoperators instead of matrices to make the notation more concise and emphasize that the results do not depend on the choice of the basis. The matrices $\MG$, $\MA$, and the other two-index matrices are treated as ordinary operators (paired indices $iα$ are treated as one index). 

The correlation matrix $\mathcal{C}_{\smash{iαjβ;kl}}$ carries four indices and can be naturally interpreted as a superoperator $\FMC$ mapping operators in one space to operators in another space. Specifically, for an operator $\MX$ in the bond space, we define a mapping to an operator $\MY$ in the space of degrees of freedom as
\begin{equation}
    \FMC\,\MX = \MY,
\end{equation}
where
\begin{equation}
    \sum_{k,l}\mathcal{C}_{iαjβ;kl}X_{kl}=Y_{iα,jβ}.
\end{equation}
Thus, the four-index correlation matrix $\mathcal{C}$ specifies the linear map from the matrix elements of $\MX$ to those of $\MY$.

In the present paper, ordinary operators are denoted by a hat ($\M X$), whereas superoperators are represented by calligraphic letters with a check (a reversed hat, $\FMC$). This representation greatly simplifies subsequent operator compositions and eigenvalue analysis.

In the superoperator representation, for any fixed (nonrandom) operator $\MX$, we have \cite{
    Erdos-matrix-dyson-equation-2019}
\begin{equation}
    \label{eq:Kraus}
    \FMC\thin\MX = \big{⟨} \MA\,\MX\MA^\dag\big{⟩}_{\!\MA},
\end{equation}
where the subscript emphasizes that the statistical averaging is taken exclusively over the random operator $\MA$.

Maps of the form (\ref{eq:Kraus}) are completely positive linear maps
and are standard in quantum information theory~\cite{
    Nielsen-quantum-computation-quantum-2010,
    Choi-completely-positive-linear-1975}. In this
analogy, the realizations of $\MA$ play the role of Kraus operators.
Unlike a quantum channel, however, the map considered here is not
required to be trace preserving, and no corresponding normalization
condition is imposed.

The properties of superoperators are well known in the theory of open quantum systems~\cite{Breuer-theory-open-quantum-2009}. To define the Hermitian conjugation of a superoperator, we use the Hilbert--Schmidt inner product in both operator spaces~\cite{Breuer-theory-open-quantum-2009},
\begin{equation}
    \bigl(\MX, \MY\bigr)_{\rm HS} \equiv \Tr{\M X^\dag\M Y},  \label{eq:scalar}
\end{equation}
where $\operatorname{Tr}[·]$ denotes the operator trace. Using this scalar product, the Hermitian conjugate $\FMC^\dag$ of a superoperator $\FMC$ is defined via
\begin{equation}
    \bigl(\MY, \FMC \MX\bigr)_{\rm HS} = \bigl(\FMC^\dag \MY, \MX\bigr)_{\rm HS},
\end{equation}
which must hold for all operators $\M X$ and $\M Y$. It follows that the action of the conjugated superoperator $\FMC^\dag$ is given by
\begin{equation}
    \FMC^\dag \MY = \big{⟨}\MA^\dag \MY \MA\big{⟩}_{\!\MA},
\end{equation}
and the matrix elements are
\begin{equation}
    \bigl(\FMC^\dag\bigr)_{kl;iαjβ} = \mathcal{C}^*_{iαjβ;kl}.  \label{eq:hc}
\end{equation}
Thus $\FMC$ maps operators in the bond space to operators
in the space of degrees of freedom, while its  Hermitian conjugate $\FMC^\dag$ maps in the opposite direction.

\subsection{Dyson equations}

The Green function $\MG$ defined by Eq.~(\ref{eq:G}) with the random
force-constant matrix $\M{Φ}$ defined by Eq.~(\ref{eq:MPhi}) satisfies
the following set of Dyson equations when the parametrically smaller
nonplanar diagrams are neglected (see Appendix~\ref{app:Dyson}):
\begin{spalign}
    \label{eq:Dyson}
    \MG &= \big(\FMC\,\MGb + \MW\big)^{-1}, 
    \\
    \MGb &= \big(\FMC^\dag\MG + \M1\big)^{-1},
\end{spalign}
where we introduce the nonrandom matrix
\begin{equation}
    \MW = \M{Φ}_0 - \M m ω^2
\end{equation}
and the secondary Green function
\begin{equation}
    \MGb = \avgs{\big(\MA^\dag\MW^{-1}\MA +\M1\big)^{-1}}_{\!\MA}.
\end{equation}
We do not take into account the mass disorder in $\M m$, since we study the quasistatic response at $ω→0$.

The operator $\MG$ acts in the space of degrees of freedom and therefore has dimensions $N_\dof \times N_\dof$, while the operator $\MGb$ acts in the auxiliary bond space and has dimensions $N_\bb \times N_\bb$. Here and below, the subscript (or superscript) {\small $\bb$} denotes operators or superoperators acting in the auxiliary space of generalized bonds rather than in the space of physical degrees of freedom. The Dyson equations can also be viewed as the saddle-point equations of a functional $ℱ(\MG,\MGb)$ given in Appendix~\ref{app:minimum}.

By substituting the Dyson equations (\ref{eq:Dyson}) into the definition of the effective force-constant matrix $\M{Φ}^\eff$ given in Eq.~(\ref{eq:Phi_eff_def}), we arrive at
\begin{equation}
    \M{Φ}^\eff = \MG^{-1} + \M m ω^2 = \M{Φ}_0 + \FMC\thin\MGb.
\end{equation}
The first term $\M{Φ}_0$ is the direct contribution present even in
the absence of disorder ($\FMC=0$), whereas the second term
$\FMC\,\MGb$ is the self-consistent disorder-induced contribution that follows from the solution of the Dyson equations (\ref{eq:Dyson}).

\subsection{Perturbation}

The introduction of inclusions or the presence of interfaces within an amorphous medium can be modeled as a perturbation of the force-constant matrix, denoted by $δ\M{Φ}$. We perform an ensemble average over structural disorder while keeping the geometry and physical properties of the inclusions and interfaces fixed. We assume that the perturbation $δ\M{Φ}$ is deterministic, while the statistical properties of the disorder, encoded in $\FMC$, remain unchanged.

The perturbed Dyson equations then read
\begin{spalign}
    \label{eq:Dyson_pert}
    \MG + δ\MG &= \big(\FMC\,(\MGb + δ\MGb)+ \MW + δ\M{Φ}\big)^{-1},  
    \\
    \MGb + δ\MGb &= \big(\FMC^\dag(\MG + δ\MG) + \M1\big)^{-1}.
\end{spalign}
These equations can be written in a more convenient form
\begin{spgather}
    \label{eq:Dyson_pert3}
    δ\MG + \big(\MG + δ\MG\big)\big(\FMC\,δ\MGb + δ\M{Φ}\big)\MG = 0,
    \\[0.5ex]
    δ\MGb + \big(\MGb + δ\MGb\big)\big(\FMC^\dag δ\MG\big)\MGb = 0.
\end{spgather}
This system of equations is algebraically exact for an arbitrary
perturbation within the planar approximation used to derive the
Dyson equations~(\ref{eq:Dyson}).

The resulting deviation of the effective force-constant matrix is
\begin{equation}
    \label{eq:dPhi_eff}
    δ\M{Φ}^\eff = \bigl(\MG + δ\MG\bigr)^{-1} - \MG^{-1} = δ\M{Φ} + \FMC\thinδ\MGb.
\end{equation}
The second term $\FMC\,δ\MGb$ represents the disorder-mediated
correction to the effective response. Since $δ\MGb$ is determined
self-consistently by Eq.~\eqref{eq:Dyson_pert3}, this term generates
a generally nonlocal dependence of $δ\M{Φ}^\eff$ on the perturbation
$δ\M{Φ}$.

\section{Small perturbation}
\label{sec:small}

In this section, we restrict our attention to the regime of small perturbations $δ\M{Φ}$. A representative example is the incorporation of nanoscale inclusions whose elastic stiffness does not differ significantly from that of the surrounding matrix. The case of a general perturbation $δ\M{Φ}$, which may be large in the region directly affected by an inclusion or interface, will be examined in the subsequent section.

\begin{figure*}
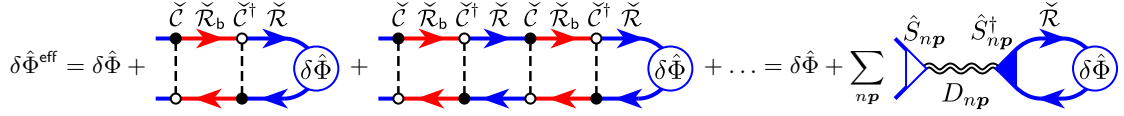

    \begin{equation*}
        δ\M{Φ}^\eff = δ\M{Φ} + \newdiag[-14pt]{47} + \newdiag[-14pt]{49}  + \ldots = δ\M{Φ} + \sum_{n\Vp} \newdiag[-14pt]{50}
    \end{equation*}
    \caption{Graphical presentation of the ladder series given by Eq.~(\ref{eq:Phi_eff_series}) and its factorized presentation using eigenoperator decomposition (\ref{eq:C_Bloch}). See Appendices \ref{app:Dyson} and \ref{app:eigval} for a more detailed explanation of the diagrams. }
    \label{fig:ladder}
\end{figure*}

Assuming that the Green-function perturbations $δ\MG$ and $δ\MGb$ are also small, Eq.~(\ref{eq:Dyson_pert3}) yields, to leading order,
\begin{spalign}
    \label{eq:Dyson_delta}
    δ\MG &= -\MG \big(\FMC\,δ\MGb + δ\M{Φ}\big)\MG,
    \\
    δ\MGb &= -\MGb \big(\FMC^\dagδ\MG\big)\MGb.
\end{spalign}
Even for a small perturbation $δ\M{Φ}$, however, the resulting
$δ\MG$ and $δ\MGb$ need not remain small in the near-critical regime, where the response is strongly amplified. This regime will be considered in the next section. 

We introduce the propagation superoperators $\FMR$ and $\FMRb$ such that
\begin{spalign}
    \FMR\,\MY &= \MG\,\MY \MG, \\
    \FMRb\thin\MX &= \MGb \MX \MGb,
\end{spalign}
for arbitrary operators $\MX$ and $\MY$. Thus, Eq.~(\ref{eq:Dyson_delta}) reads as
\begin{spalign}
    \label{eq:Dyson_delta_R}
    δ\MG &= - \FMR \big(\FMC\,δ\MGb + δ\M{Φ}\big),
    \\
    δ\MGb &= - \FMRb\thin\FMC^\dagδ\MG.
\end{spalign}
The solution of this linear system gives $δ\MG$ and $δ\MGb$. Therefore, the resulting correction to the effective force-constant matrix given by Eq.~(\ref{eq:dPhi_eff}) has the following series representation:
\begin{equation}
    δ\M{Φ}^\eff = δ\M{Φ} + \FMC \FMRb \FMC^\dag \FMR\,δ\M{Φ} + \FMC \FMRb \FMC^\dag \FMR\,\FMC \FMRb \FMC^\dag \FMR \, δ\M{Φ} + \ldots
    \label{eq:Phi_eff_series}
\end{equation}
It has a familiar ladder form, presented in Fig.~\ref{fig:ladder}  (see Appendices \ref{app:Dyson} and \ref{app:eigval} for a more detailed explanation of the diagram technique).

\subsection{Eigenvalue analysis and Bloch's theorem}

To analyze the linear system~\eqref{eq:Dyson_delta_R} and construct
a natural basis that will also be useful for finite perturbations,
we consider the associated eigenvalue problem.  In what follows, we examine the zero-frequency limit by parameterizing the frequency as $\omega = i\zeta$ with $ζ>0$ and subsequently taking the limit $\zeta \to 0^+$ \cite{Conyuh-largescale-exponential-correlations-2026}. In this case, Green functions $\MG$ and $\MGb$ are Hermitian, and the corresponding propagation superoperators are, therefore, also Hermitian:
\begin{equation}
    \FMR^\dag = \FMR, \qquad \FMRb^\dag = \FMRb,
    \label{eq:R_Herm}
\end{equation}
This allows us to define the set of nonzero eigenvalues $σ_v$ and the eigenoperators $\MS_v$ and $\MSb_v$ such that (see Appendix \ref{app:eigval})
\begin{spalign}
    \label{eq:S_eigval}
    \FMC\,\FMRb\thin\MSb_v &= σ_v \MS_v,
    \\
    \FMC^\dag\FMR\,\MS_v &= σ_v \MSb_v
\end{spalign}
with orthogonality relations
\begin{spalign}
    \label{eq:orthonorm}
    \Tr{\MS^\dag_v\,&\FMR\,\MS_{v'}} = δ_{vv'},
    \\
    \Tr{\MSb^\dag_v\,&\FMRb\,\MSb_{v'}} = δ_{vv'}.
\end{spalign}
The eigenvalues $σ_v$ are real and satisfy $0<σ_v\leq1$ (see Appendices~\ref{app:eigval} and~\ref{app:minimum}). Each index $v$ labels a disorder-mediated response channel, specified by a pair of eigenoperators $\MS_v$ and $\MSb_v$ coupled through the nonzero eigenvalue $σ_v$. If the largest eigenvalue reaches unity, the corresponding channel becomes critical and the system is referred to as critical.

\begin{figure}
    \centering
    \includegraphics[scale=0.7]{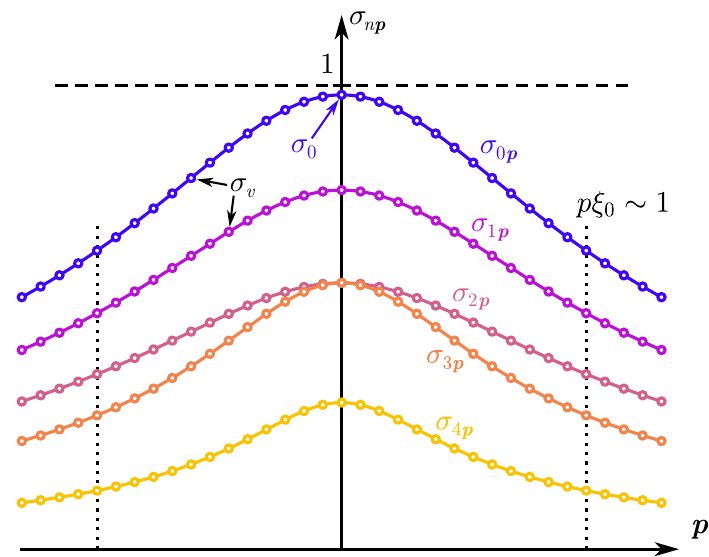}
    \caption{Schematic illustration of the branches $σ_{n\Vp}$ (colored lines) and discrete eigenvalues $σ_v$ for the finite system (colored points) for $σ_0 = 0.98$. The vertical dotted lines show the position of $pξ_0\sim 1$.}
    \label{fig:branches}
\end{figure}

For a statistically homogeneous unperturbed medium without boundaries, disorder averaging restores translational invariance on the reference lattice. The corresponding operators and superoperators therefore commute with lattice translations, so that Bloch's theorem can be applied (see Appendix~\ref{app:Bloch}). The channel index $v$ can then be written as a pair $(n,\Vp)$, where $\Vp$ is the Bloch wavevector and $n$ labels the branch. Thus, each pair $(n,\Vp)$ specifies a disorder-mediated response channel, while the set of channels with fixed $n$ and varying $\Vp$ forms a branch of eigenvalues $σ_{n\Vp}$, as illustrated in Fig.~\ref{fig:branches}. We refer to $n=0$ as the upper branch.

In Bloch's notation, Eq.~(\ref{eq:S_eigval}) reads as
\begin{spalign}
    \label{eq:S_eigval_Bloch}
    \FMC\,\FMRb\thin\MSb_{n\Vp} &= σ_{n\Vp} \MS_{n\Vp},
    \\
    \FMC^\dag\FMR\,\MS_{n\Vp} &= σ_{n\Vp} \MSb_{n\Vp},
\end{spalign}
where the eigenoperators $\MS_{n\Vp}$ and $\MSb_{n\Vp}$ satisfy the orthogonality relations
\begin{spalign}
    \label{eq:orthog}
    \Tr{\MS^\dag_{n\Vp}\,\FMR\,\MS_{n'\Vp'}} &= δ_{nn'}δ_{\Vp\Vp'},
    \\
    \Tr{\MSb^\dag_{n\Vp}\,\FMRb\,\MSb_{n'\Vp'}} &= δ_{nn'}δ_{\Vp\Vp'}.
\end{spalign}
With this labeling, the correlation superoperator can be factorized as
\begin{spalign}
    \label{eq:C_Bloch}
    \FMC\thin\MX &= \sum_{n\Vp} \MS_{n\Vp}\, σ_{n\Vp} \Tr{\MSb_{n\Vp}^\dag \MX},
    \\
    \FMC^\dag \MY &= \sum_{n\Vp} \MSb_{n\Vp}\, σ_{n\Vp} \Tr{\MS_{n\Vp}^\dag \MY},
\end{spalign}
where $\MX$ and $\MY$ are arbitrary operators. In the thermodynamic limit $N \to \infty$, the discrete summation over $\Vp$ in Eq.~(\ref{eq:C_Bloch}) is replaced by integration over the first Brillouin zone of the reference lattice.

Hermitian conjugation of eigenoperators $\MS_{n\Vp}$ and $\MSb_{n\Vp}$ reverses the direction of the wavevectors (see Appendix~\ref{app:Herm})
\begin{spalign}
    \label{eq:Snp_Herm}
    \MS^\dag_{n\Vp} &= \MS_{n,-\Vp}, 
    \\
    \MSb^\dag_{n\Vp} &= \MSb_{n,-\Vp}.
\end{spalign}

\subsection{Small perturbation in the eigenoperator basis}

The perturbation $δ\M{Φ}$ can be expressed as an expansion in the basis of eigenoperators $\MS_{n\Vp}$:
\begin{equation}
    \label{eq:Phi_expansion}
    δ\M{Φ} = \sum_{n\Vp} δφ_{n\Vp}\,\MS_{n\Vp} + δ\M{Φ}_\perp,
\end{equation}
where the expansion coefficients are given by
\begin{equation}
    δφ_{n\Vp} = \Tr{\MS_{n\Vp}^\dag\thin\FMR\,δ\M{Φ}}.
\end{equation}
The presentation (\ref{eq:Phi_expansion}) explicitly involves contributions from all wavevectors $\Vp$, reflecting the fact that the perturbation $δ\M{Φ}$ is, in general, spatially inhomogeneous. The additional term $δ\M{Φ}_\perp$ represents the remainder that is orthogonal to all basis operators, such that $\Tr{\MS_{n\Vp}^\dag\thin\FMR\,δ\M{Φ}_\perp}=0$ for all $n$ and $\Vp$. The presence of such an additional term arises from the fact that the basis operators do not form a complete set (see Appendix~\ref{app:eigval}).

Multiplying Eq.~(\ref{eq:Dyson_delta_R}) by $\MS_{n\Vp}^\dag$ and $\MSb_{n\Vp}^\dag$ and taking the trace, we obtain
\begin{spalign}
    \label{eq:delta_ab}
    &δa_{n\Vp} + σ_{n\Vp}\,δb_{n\Vp} + δφ_{n\Vp} = 0,
    \\
    &δb_{n\Vp} + σ_{n\Vp}\,δa_{n\Vp} = 0,
\end{spalign}
where the coefficients $δa_{n\Vp}$ and $δb_{n\Vp}$ denote the projections of the Green-function perturbations onto the basis operators
\begin{spalign}
    δa_{n\Vp} &= \Tr{\MS_{n\Vp}^\dag\thinδ\MG},
    \\
    δb_{n\Vp} &= \Tr{\MSb_{n\Vp}^\dag\thinδ\MGb}.
\end{spalign}
From Eq.~(\ref{eq:delta_ab}), we obtain the coefficients explicitly:
\begin{spalign}
    \label{eq:small_resp_p}
    δa_{n\Vp} &= - \frac{δφ_{n\Vp}}{1 - σ^2_{n\Vp}},
    \\
    δb_{n\Vp} &= \frac{σ_{n\Vp}\,δφ_{n\Vp}}{1 - σ^2_{n\Vp}}.
\end{spalign}
The resulting perturbation of the Green functions is
\begin{spalign}
    δ\MG &= \FMR\sum_{n\Vp} δa_{n\Vp}\,\MS_{n\Vp} - \FMR\,δ\M{Φ}_\perp,
    \\
    δ\MGb &= \FMRb\sum_{n\Vp} δb_{n\Vp}\,\MSb_{n\Vp},
\end{spalign}
where the orthogonal remainder $-\FMR\,δ\M{Φ}_\perp$ follows from Eq.~(\ref{eq:Dyson_delta_R}). Putting the resulting $δ\MGb$ into Eq.~(\ref{eq:dPhi_eff}), we obtain the change in the effective force-constant matrix
\begin{equation}
    δ\M{Φ}^\eff = δ\M{Φ} + \sum_{n\Vp}\MS_{n\Vp}D_{n\Vp}δφ_{n\Vp},
\end{equation}
where we have introduced the disorder-induced kernel
\begin{equation}
    D_{n\Vp} = \frac{σ^2_{n\Vp}}{1-σ^2_{n\Vp}}.
\end{equation}
The graphical presentation of this result is shown on the right side of Fig.~\ref{fig:ladder}. 

The most important contribution comes from the upper branch $n=0$, which has the smallest denominator $1-σ^2_{n\Vp}$ at $\Vp=0$. According to the Perron--Frobenius theorem, the upper branch is nondegenerate at $\Vp=0$~\cite{Conyuh-largescale-exponential-correlations-2026}. For an isotropic disorder-averaged medium, this nondegenerate branch can depend on the wavevector only through $p^2$ to leading order near $\Vp=0$. Therefore,
\begin{equation}
    \label{eq:sigma2_expansion}
    σ_{0\Vp}^2 = σ_0^2 - ξ_0^2p^2 + \mathcal{O}(p^4),
\end{equation}
where $ξ_0$ is a typical structural length scale in the system, expected to be close to the interatomic distance in amorphous solids. 
For $1 - σ_0 \ll 1$ and $|\Vp|\ll1/ξ_0$, the leading
near-critical contribution is
\begin{equation}
    \label{eq:D_0p}
    D_{0\Vp} ≃ \frac{σ_0^2}{1-σ_0^2}\frac{1}{1 + ξ^2 p^2},
\end{equation}
where
\begin{equation}
    ξ = ξ_0/\sqrt{1-σ_0^2} \gg ξ_0
\end{equation} 
is a nontrivial length scale that can significantly exceed the structural length scale $ξ_0$. The limit $ξ\ggξ_0$ defines the regime of \emph{strong disorder}, or, more technically, the \emph{near-critical regime}.

The kernel $D_{0\Vp}$ can be viewed as a zero-frequency diffusion-relaxation kernel with a diffusion length $ξ$ or a screened Poisson kernel with a screening length $ξ$. In the elastic problem, $ξ$ corresponds to the nonaffine length, which controls both the long-range correlations of nonaffine displacements~\cite{Conyuh-largescale-exponential-correlations-2026} and the spatial propagation of boundary-induced constraints, as discussed in Section~\ref{sec:qualitative}.

\subsection{Real-space presentation and Wannier operators}

For further analysis, it is more convenient to work in real space than in reciprocal space. By analogy with the well-known Wannier functions~\cite{Wannier-structure-electronic-excitation-1937, Marzari-maximally-localized-wannier-2012}, we introduce the Wannier operators
\begin{spalign}
    \MS_{n\Vr} &≡ \frac{1}{\sqrt{N}}\sum_{\Vp} e^{-i\Vp\Vr}\MS_{n\Vp},
    \\
    \MSb_{n\Vr} &≡ \frac{1}{\sqrt{N}}\sum_{\Vp} e^{-i\Vp\Vr}\MSb_{n\Vp}.
\end{spalign}
Here $\Vr$ labels the center of the Wannier operator. For the
short-ranged homogeneous system considered here, the Wannier
operators can be chosen such that their matrix elements are localized
around $\Vr$ on the microscopic scale $ξ_0\ll ξ$.

The Wannier operators are mutually orthogonal,
\begin{spalign}
    \Tr{\MS^\dag_{n\Vr}\,&\FMR\,\MS_{n'\Vr'}} = δ_{nn'}δ_{\Vr\Vr'},
    \\
    \Tr{\MSb^\dag_{n\Vr}\,&\FMRb\,\MSb_{n'\Vr'}} = δ_{nn'}δ_{\Vr\Vr'}
\end{spalign}
and are Hermitian:
\begin{spalign}
    \MS^\dag_{n\Vr} &= \MS_{n\Vr},
    \\
    \MSb^\dag_{n\Vr} &= \MSb_{n\Vr}.
\end{spalign}

Now, the perturbation of the force-constant matrix can be decomposed into the Wannier operators as
\begin{equation}
    δ\M{Φ} = \sum_{n\Vr} δφ_{n\Vr} \MS_{n\Vr} + δ\M{Φ}_\perp
\end{equation}
with the expansion coefficients given by
\begin{equation}
    δφ_{n\Vr} = \Tr{\MS_{n\Vr}^\dag\,\FMR\,δ\M{Φ}} = \frac{1}{\sqrt{N}}\sum_{\Vp}δφ_{n\Vp}e^{i\Vp\Vr}.
\end{equation}
The resulting perturbation of the Green functions is
\begin{spalign}
    \label{eq:Wannier}
    δ\MG &= \FMR\sum_{n\Vr} δa_{n\Vr}\,\MS_{n\Vr} - \FMR\,δ\M{Φ}_\perp,
    \\
    δ\MGb &= \FMRb\sum_{n\Vr} δb_{n\Vr}\,\MSb_{n\Vr},
\end{spalign}
where the coefficients are obtained from the Fourier transform as follows
\begin{spalign}
    δa_{n\Vr} &= \frac{1}{\sqrt{N}}\sum_{\Vp}δa_{n\Vp}e^{i\Vp\Vr},\\
    δb_{n\Vr} &= \frac{1}{\sqrt{N}}\sum_{\Vp}δb_{n\Vp}e^{i\Vp\Vr}.
\end{spalign}

As a result, the effective force-constant matrix $δ\M{Φ}^\eff$ has the following real-space presentation
\begin{equation}
    δ\M{Φ}^\eff = δ\M{Φ} + \sum_{n\Vr\Vr'}\MS_{n\Vr} D_n(\Vr-\Vr')δφ_{n\Vr'},
\end{equation}
where $D_n(\Vr)$ is a real-space disorder-induced kernel
\begin{equation}
    D_n(\Vr) = \frac{1}{N}\sum_{\Vp}D_{n\Vp} e^{i\Vp\Vr}.
\end{equation}

As stated in the previous subsection, the long-wavelength response is
dominated by the upper branch $n=0$. Neglecting the lower branches
with $n>0$, the dominant long-wavelength part of the effective
force-constant perturbation can be written as
\begin{equation}
    δ\M{Φ}^\eff ≃ \sum_{\Vr} a α(\Vr) \MS_{0\Vr},
\end{equation}
where $a$ is a normalization constant and $α(\Vr)$ is a dimensionless contrast given by the solution to the screened Poisson equation
\begin{equation}
    \label{eq:Poisson}
    α(\Vr) - ξ^2Δα(\Vr) = φ(\Vr),
\end{equation}
where
\begin{equation}
    \label{eq:ext_pert}
    φ(\Vr) = \frac{δφ_{0\Vr}}{a(1-σ_0^2)}
\end{equation}
is the normalized external perturbation. The physically relevant value of the constant $a$ will be determined in the next section, where the case of a finite perturbation is analyzed.

We can consider $α(\Vr)$ as a continuous function because $ξ\gg ξ_0$. Accordingly, such a separation of scales also allows us to define spatial variations of elastic moduli. In particular, the spatial variation of the bulk modulus $\KK(\Vr)$ and the shear modulus $\GG(\Vr)$ (not to be confused with the Green function $\MG$) takes the following form:
\begin{spalign}
    \label{eq:moduli}
    \KK(\Vr) = \KK_0 + \KK_1 α(\Vr),
    \\
    \GG(\Vr) = \GG_0 + \GG_1 α(\Vr),
\end{spalign}
where $\KK_0$ and $\GG_0$ are the elastic moduli of the infinite amorphous medium, and $\KK_1$ and $\GG_1$ are constants that depend on the disorder strength. Therefore, the function $α(\Vr)$ can be viewed as a \emph{disorder-induced elastic contrast}. The elastic moduli $\KK(\Vr)$ and $\GG(\Vr)$ in Eq.~(\ref{eq:moduli}) are treated as the effective ones that represent the average response of the disordered elastic medium. 

Thus, although the microscopic elastic response is tensorial, its
dominant long-wavelength spatial variation is governed by the single
scalar field $α(\Vr)$. The tensorial information is retained in the
eigenoperator $\MS_{0\Vr}$ and, for an isotropic elastic medium, in
the amplitudes $\KK_1$ and $\GG_1$, whereas the spatial profile is
common to both elastic moduli.

\section{Finite perturbation}
\label{sec:finite}

According to Eq.~(\ref{eq:D_0p}), the response associated with the upper branch is strongly enhanced as $σ_0\to1$. Consequently, even a small perturbation $δ\M{Φ}$ can produce large variations of the Green functions and of the effective force-constant matrix, so that the linearized treatment of Sec.~\ref{sec:small} is no longer sufficient. We therefore return to the nonlinear Dyson equations (\ref{eq:Dyson_pert3}), which do not require $δ\M{Φ}$ to be small everywhere. In particular, $δ\M{Φ}$ may be large inside an inclusion or within a microscopic interfacial region.

\begin{figure}
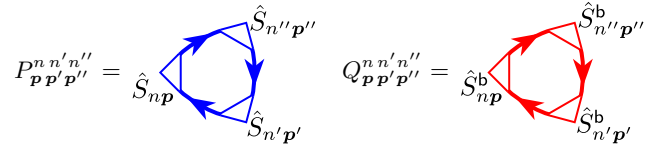

    \begin{equation*}
        P^{n\,n'n''}_{\Vp\,\Vp'\Vp''} = \newdiag[-25pt]{51}
        \quad 
        Q^{n\,n'n''}_{\Vp\,\Vp'\Vp''} = \newdiag[-25pt]{52}
    \end{equation*}
    \caption{Graphical presentation of the nonlinear coefficients given by Eq.~(\ref{eq:PQ}).}
    \label{fig:PQ}
\end{figure}

In the following analysis, we focus on the disorder-mediated part of the response and neglect components orthogonal to the disorder-mediated subspaces introduced in Appendix~\ref{app:eigval}. Projecting Eqs.~(\ref{eq:Dyson_pert3}) onto the corresponding eigenoperators $\MS_{n\Vp}$ and $\MSb_{n\Vp}$ and taking the trace, we obtain
\begin{spalign}
    \label{eq:nonlin_gen}
    & δa_{n\Vp} + δφ_{n\Vp}^\eff
    + \!\sum_{n'n''\Vp'\Vp''}\!\!\!P^{\,n\,n'n''}_{-\Vp,\Vp',\Vp''}\thinδa_{n'\Vp'}\thinδφ_{n''\Vp''}^\eff = 0,\!
    \\
    &δb_{n\Vp} + δ\tilde{a}_{n\Vp}
    + \!\sum_{n'n''\Vp'\Vp''}\!\! Q^{\,n\,n'n''}_{-\Vp,\Vp',\Vp''}\thinδb_{n'\Vp'}\thinδ\tilde{a}_{n''\Vp''} = 0,
\end{spalign}
where
\begin{spalign}
    δ\tilde{a}_{n\Vp} &= σ_{n\Vp}δa_{n\Vp},
    \\
    δφ_{n\Vp}^\eff &= δφ_{n\Vp} + σ_{n\Vp}\thinδb_{n\Vp},
\end{spalign}
and we introduced the following nonlinear coefficients:
\begin{spalign}
    \label{eq:PQ}
    P^{n\,n'n''}_{\Vp\,\Vp'\Vp''} &= \Tr{\MG\,\MS_{n\Vp}\,\MG\,\MS_{n'\Vp'}\,\MG\,\MS_{n''\Vp''}},
    \\
    Q^{n\,n'n''}_{\Vp\,\Vp'\Vp''} & = \Tr{\MGb\,\MSb_{n\Vp}\,\MGb\,\MSb_{n'\Vp'}\,\MGb\,\MSb_{n''\Vp''}}.
\end{spalign}
These coefficients are nonzero only if the momentum-conservation condition $\Vp+\Vp'+\Vp''=0$ is fulfilled, and they are invariant under cyclic permutations of the index pairs $(n,\Vp)$, $(n',\Vp')$, and $(n'',\Vp'')$. The graphical presentation of these coefficients is given in Fig.~\ref{fig:PQ}.

The resulting perturbation of the effective force-constant matrix is given by the coefficients $δφ_{n\Vp}^\eff$:
\begin{equation}
    δ\M{Φ}^\eff = \sum_{n\Vp}δφ_{n\Vp}^\eff\MS_{n\Vp}.
\end{equation}

To proceed with the analytical treatment, we introduce several assumptions. We focus on regions where the direct perturbation $δ\M{Φ}$ is small and varies smoothly, whereas the induced response need not be small. A particularly important situation arises in the presence of inclusions or boundaries. In this case, the perturbation and the resulting $δ\M{Φ}^\eff$ may be large inside the inclusion or within a microscopic interfacial region. Our interest, however, is in the long-range response of the surrounding amorphous medium, where the direct structural modification represented by $δ\M{Φ}$ is small or absent.

Accordingly, within the region of interest we assume that $δ\M{Φ}$ and its expansion coefficients $δφ_{0\Vr}$ are small in magnitude and vary smoothly. At the same time, the variations of the Green functions $δ\MG$ and $δ\MGb$ may attain large amplitudes. Their spatial variation, however, occurs on the collective length scale $ξ$, which is much larger than the microscopic scale in the near-critical regime. Spatial derivatives are therefore correspondingly small and can be treated perturbatively.

\subsection{Strongly disordered regime}

Let us assume that $σ_0\equiv σ_{0,\Vp=0}$ is close to unity.
Equation~(\ref{eq:small_resp_p}) then shows that the response to a
weak perturbation is strongly enhanced, so that the upper branch
$n=0$ is near critical. We assume that this is the only near-critical
branch, while the branches with $n>0$ remain separated from unity. This means that $δa_{0\Vp}$ and $δb_{0\Vp}$ are much larger than $δφ_{0\Vp}$ and much larger than any $δa_{n\Vp}$ and $δb_{n\Vp}$ for $n>0$. 

This regime corresponds to a strongly disordered amorphous solid. In this case, the coefficients $δa_{0\Vp}$ and $δb_{0\Vp}$ play a major role and can be large. In the linear regime, we have seen that $δa_{0\Vp}$ and $δb_{0\Vp}$ are small for $|\Vp| \gg 1/ξ$. This implies that the corresponding real-space variation is smooth, with a characteristic length scale $ξ$, which diverges as $σ_0\to1$. Therefore, we assume that the typical length scale of $δa_{0\Vp}$ and $δb_{0\Vp}$ remains much larger than the atomic scale in the nonlinear regime as well.

In the limit $σ_0 → 1$, the nonlinear mixing coefficients $P^{n00}_{000}$ and $Q^{n00}_{000}$ with $n>0$ vanish (see Appendix \ref{app:mixing}). Because the wavevector $\Vp$ appearing in $δa_{0\Vp}$ and $δb_{0\Vp}$ is small and of order $1/ξ$, we can neglect the net influence of the near-critical upper branch on the remaining branches, which is mediated by the following terms:
\begin{spgather}
    \sum_{\Vp'\Vp''}P^{\,n\,0\,0}_{-\Vp,\Vp',\Vp''}\thinδa_{0\Vp'}\thinδφ_{0\Vp''}^\eff,
    \\
    \sum_{\Vp'\Vp''}Q^{\,n\,0\,0}_{-\Vp,\Vp',\Vp''}\thinδb_{0\Vp'}\thinδ\tilde{a}_{0\Vp''}
\end{spgather}
for $n>0$. Specifically, the difference between $P^{\,n\,0\,0}_{-\Vp,\Vp',\Vp''}$ evaluated at $|\Vp'|\sim |\Vp''|\sim 1/ξ$ and its value at $\Vp'=\Vp''=0$ is negligible. An analogous argument applies to $Q^{\,n\,0\,0}_{-\Vp,\Vp',\Vp''}$.

Therefore, to leading order in the near-critical long-wavelength
regime, we neglect the nonlinear mixing with the remaining branches
and retain only the upper branch. The equations for the upper branch in real space are
\begin{spalign}
    &δa_{0\Vr} + δφ^\eff_{0\Vr} +\sum_{\Vr'\Vr''} P_{\Vr\Vr'\Vr''}δa_{0\Vr'}δφ^\eff_{0\Vr''} = 0,  \label{eq:nonlin_crit_1}
    \\
    &δb_{0\Vr} + δ\tilde{a}_{0\Vr}
     + \sum_{\Vr'\Vr''}Q_{\Vr\Vr'\Vr''}δb_{0\Vr'}δ\tilde{a}_{0\Vr''} = 0
\end{spalign}
with
\begin{spalign}
    \label{eq:ab_tilde}
    δ\tilde{a}_{0\Vr} &= \sum_{\Vr'} σ_{0,\Vr-\Vr'}δa_{0\Vr'},
    \\
    δφ^\eff_{0\Vr} &= δφ_{0\Vr} + \sum_{\Vr'} σ_{0,\Vr-\Vr'}δb_{0\Vr'},
\end{spalign}
and the nonlinear coupling coefficients
\begin{spalign}
    \label{eq:PQ_0_r}
    P_{\Vr\Vr'\Vr''} &= \Tr{\MG\,\MS_{0\Vr}\,\MG\,\MS_{0\Vr'}\,\MG\,\MS_{0\Vr''}},
    \\
    Q_{\Vr\Vr'\Vr''} & = \Tr{\MGb\,\MSb_{0\Vr}\,\MGb\,\MSb_{0\Vr'}\,\MGb\,\MSb_{0\Vr''}}.
\end{spalign}

In real space, the nonlinear kernels are difference kernels such that $P_{\Vr\Vr'\Vr''} = P_{0,\Vr'-\Vr,\Vr''-\Vr}$ and $Q_{\Vr\Vr'\Vr''} = Q_{0,\Vr'-\Vr,\Vr''-\Vr}$. 

For smooth functions $δa_{0\Vr}$ and $δb_{0\Vr}$, Eqs.~(\ref{eq:ab_tilde}) and (\ref{eq:sigma2_expansion}) give
\begin{spalign}
    \label{eq:tilde_ab_smooth}
    δ\tilde{a}_{0\Vr} &= σ_0δa_{0\Vr} + \frac{ξ_0^2}{2σ_0}Δδa_{0\Vr},
    \\
    δφ^\eff_{0\Vr} &= δφ_{0\Vr} + σ_0δb_{0\Vr} + \frac{ξ_0^2}{2σ_0}Δδb_{0\Vr}.
\end{spalign}

For two smooth functions $f(\Vr)$ and $g(\Vr)$, the nonlinear kernels give (see Appendix~\ref{app:nl_kernel})
\begin{spalign}
    \label{eq:nl_kernel_exp}
    \sum_{\Vr'\Vr''}P_{\Vr\Vr'\Vr''}f(\Vr') g(\Vr'') &= \frac{\bigl[f g + ν_\aa \mathcal{D}(f, g)\bigr]_\Vr}{a},
    \\
    \sum_{\Vr'\Vr''}Q_{\Vr\Vr'\Vr''}f(\Vr') g(\Vr'') &= \frac{\bigl[f g + ν_\bb \mathcal{D}(f, g)\bigr]_\Vr}{b},
\end{spalign}
where we introduce the second-order differential operator
\begin{equation}
    \mathcal{D}(f, g) = (∇f)·(∇g) + fΔg + gΔf.
\end{equation}
Four constants $a$, $b$, $ν_\aa$, and $ν_\bb$ depend on the zeroth and second moments of the kernels given by (\ref{eq:PQ_0_r}).

Thus
\begin{align}
    & δa_{0\Vr} + δφ^\eff_{0\Vr} + \frac{1}{a} δa_{0\Vr}δφ^\eff_{0\Vr} + \frac{ν_\aa}{a}\mathcal{D}(δa_{0\Vr}, δφ^\eff_{0\Vr}) = 0,
    \\
    &δb_{0\Vr} + δ\tilde{a}_{0\Vr} + \frac{1}{b} δb_{0\Vr}δ\tilde{a}_{0\Vr} + \frac{ν_\bb}{b}\mathcal{D}(δb_{0\Vr},δ\tilde{a}_{0\Vr}) = 0.
\end{align}
These equations can be written in a concise form as
\begin{align}
    \label{eq:abD1}
    &(a + δa_{0\Vr})(a + δφ^\eff_{0\Vr}) + ν_\aa \mathcal{D}(δa_{0\Vr}, δφ^\eff_{0\Vr}) = a^2,
    \\
    \label{eq:abD2}    
    &(b + δb_{0\Vr})(b + δ\tilde{a}_{0\Vr}) + ν_\bb \mathcal{D}(δb_{0\Vr},δ\tilde{a}_{0\Vr}) = b^2.
\end{align}
Since we consider smooth functions $δa_{0\Vr}$ and $δb_{0\Vr}$, the second-order derivatives represented by $\mathcal{D}$ are small. Thus, we can neglect other derivatives in $δ\tilde{a}_{0\Vr}$ and $δφ^\eff_{0\Vr}$. We can also neglect $δφ_{0\Vr}$ in $δφ^\eff_{0\Vr}$. Therefore, we can write the second-order derivatives as
\begin{multline}
    \mathcal{D}(δa_{0\Vr}, δφ^\eff_{0\Vr}) ≈ \mathcal{D}(δb_{0\Vr},δ\tilde{a}_{0\Vr}) \approx Ξ(\Vr) 
    \\
    = 
    σ_0[(∇δa_{0\Vr})·(∇δb_{0\Vr}) + δb_{0\Vr}Δδa_{0\Vr} + δa_{0\Vr}Δδb_{0\Vr}].
\end{multline}

We parameterize the upper-branch contribution to the effective force-constant perturbation as
\begin{equation}
    δφ^\eff_{0\Vr} = a α(\Vr),
\end{equation}
where $α(\Vr)$ is a smooth function that characterizes the elastic contrast and the coefficient $a$ is given by the expansions (\ref{eq:nl_kernel_exp}). Using Eq.~(\ref{eq:abD1}) and treating the second-order derivatives as perturbative corrections, we obtain
\begin{equation}
    \label{eq:delta_a_alpha}
    δa_{0\Vr} = -\frac{aα(\Vr)}{1+α(\Vr)} - \frac{ν_\aa Ξ(\Vr)}{a(1 + α(\Vr))}.
\end{equation}

From Eq.~(\ref{eq:tilde_ab_smooth}) we obtain, to the same order of approximation,
\begin{spalign}
    δ\tilde{a}_{0\Vr} &= σ_0\Bigl[-\frac{aα(\Vr)}{1+α(\Vr)} - \frac{ν_\aaΞ(\Vr)}{a(1+α(\Vr))}  - \frac{a ξ_0^2}{2σ_0^2}Δ\frac{α(\Vr)}{1 + α(\Vr)}\Bigr],
    \\
    δb_{0\Vr} &= \frac{1}{σ_0}\big[a α(\Vr) - \frac{a ξ_0^2}{2σ_0^2} Δ α(\Vr) - δφ_{0\Vr}\big].
\end{spalign}
Substituting these expressions into Eq.~(\ref{eq:abD2}), we get
\begin{multline}
    \Bigl[\frac{b}{σ_0}-\frac{aα(\Vr)}{1+α(\Vr)} - \frac{ν_\aaΞ(\Vr)}{a(1 + α(\Vr))}  + \frac{a ξ_0^2}{2σ_0^2}Δ\frac{1}{1 + α(\Vr)}\Bigr]
    \\
    ×\big[bσ_0 + a α(\Vr) - \frac{a ξ_0^2}{2σ_0^2} Δ α(\Vr) - δφ_{0\Vr}\big] + ν_\bbΞ(\Vr) = b^2.
\end{multline}
We retain terms only to first order in $1-σ_0$, $1-a/b$, and
second-order spatial derivatives. The difference $ν_\bb-ν_\aa$ is also
small in the near-critical regime (see Appendix~\ref{app:nl_kernel}).
Neglecting products of these small corrections and using the identity
\begin{equation}
     \frac{1}{f(\Vr)}Δf(\Vr) - f(\Vr)Δ\frac{1}{f(\Vr)} = 2 Δ \ln f(\Vr),
\end{equation}
we obtain
\begin{equation}
    \label{eq:main_res}
    η α(\Vr) + \frac{(1-η)α(\Vr) - φ(\Vr)}{1 + α(\Vr)} = ξ^2 Δ \ln[1+ α(\Vr)],
\end{equation}
where $η = (1-σ_0 a/b)/(1 - σ_0^2)$ is a constant parameter and $φ(\Vr) = δφ_{0\Vr}/[a(1-σ_0^2)]$ is the external perturbation given by Eq.~(\ref{eq:ext_pert}). The previously obtained Eq.~(\ref{eq:alpha1}) corresponds to the case $η=1$ in the unperturbed region.

It is useful to introduce the field $s(\Vr)$ such that
\begin{equation}
    α(\Vr)=e^{s(\Vr)}-1.
\end{equation}
Equation~(\ref{eq:main_res}) then takes the more symmetric form
\begin{multline}
    \label{eq:main_s}
    η\bigl(e^{s(\Vr)}-1\bigr)
    +(1-η)\bigl(1-e^{-s(\Vr)}\bigr)
    \\
    = ξ^2Δs(\Vr)+φ(\Vr)e^{-s(\Vr)}.
\end{multline}
This equation can be written as the Euler--Lagrange equation of the
reduced scalar functional
\begin{equation}
    \label{eq:F_reduced}
    ℱ_{\mathsf{red}}[s]
    =
    \int 
    \biggl\{
        \frac{ξ^2}{2}|\nabla s|^2
        +U_η(s)
        +φ(\Vr)\bigl(e^{-s}-1\bigr)
    \biggr\}\,\dd\Vr
\end{equation}
with
\begin{equation}
    U_η(s)
    =
    η(e^s-s-1)
    +(1-η)(e^{-s}+s-1).
\end{equation}
In the unperturbed region, where $φ(\Vr)=0$, the local part of this
functional is $U_η(s)$. For $0\leqη\leq1$, one has $U_η(s)\geq0$, with its minimum at $s=0$, and the reduced functional remains bounded from below. For $η<0$ or $η>1$, by contrast, $U_η(s)$ becomes unbounded from below as $s\to+\infty$ or $s\to-\infty$, respectively.
Since Eq.~(\ref{eq:main_res}) is obtained by reducing the stable
Dyson problem to the near-critical branch, we restrict the parameter
to $0 ≤ η ≤ 1$, for which the reduced description retains stable
finite-contrast solutions. Outside this range, the instability of the reduced functional indicates that contributions neglected in the single-branch approximation can no longer be disregarded.

\begin{figure}
    \centering
    \includegraphics[scale=0.8]{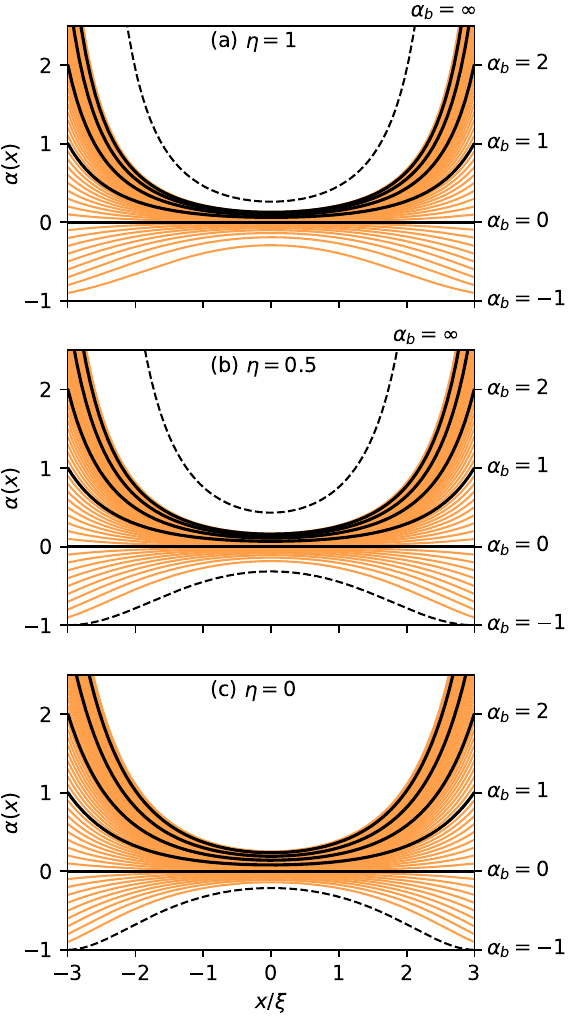}
    \caption{One-dimensional solutions of Eq.~(\ref{eq:main_res}) for different values of the parameter $η$: (a) $η = 1$, (b) $η = 0.5$, and (c) $η = 0$. Different lines correspond to different boundary values $α(±W/2) = α_b$ at the edges of the displayed region of width $W=6ξ$. Solutions with integer values of $α_b$ are represented by black lines. Dashed lines show the limiting solutions for the minimum and maximum possible boundary values, $α_b=-1$ and $α_b=\infty$, respectively, whenever such solutions exist.}
    \label{fig:alpha_x}
\end{figure}

Equation (\ref{eq:main_res}) is the main result of this paper. It represents a significant modification of the response of amorphous solids in the case of strong disorder. It generalizes the simple screened Poisson equation (\ref{eq:Poisson}) to the case of large values of the disorder-induced elastic contrast $α(\Vr)$. At the same time, the spatial variation of the bulk modulus $\KK(\Vr)$ and the shear modulus $\GG(\Vr)$ has the same form:
\begin{spalign}
    \KK(\Vr) = \KK_0 + \KK_1 α(\Vr),
    \\
    \GG(\Vr) = \GG_0 + \GG_1 α(\Vr).
\end{spalign}

The reduced equation (\ref{eq:main_s}) possesses an additional symmetry. In the
unperturbed region, $φ(\Vr)=0$, it is invariant under the simultaneous
transformations
\begin{equation}
    s(\Vr)\leftrightarrow -s(\Vr),
    \qquad
    η\leftrightarrow1-η.
\end{equation}
In terms of the elastic contrast, this corresponds to
\begin{equation}
    α(\Vr)\leftrightarrow-\frac{α(\Vr)}{1+α(\Vr)}.
\end{equation}
Note that $-α(\Vr)/[1+α(\Vr)]$ provides the leading contribution to
$δ\MG$, through the coefficients $δa_{0\Vr}$ in
Eq.~(\ref{eq:delta_a_alpha}). Thus, in the unperturbed region, the
spatial variation of the Green function is governed by an equation of
the same form as Eq.~(\ref{eq:main_res}).

\begin{figure}[t!]
    \centering
    \includegraphics[scale=0.8]{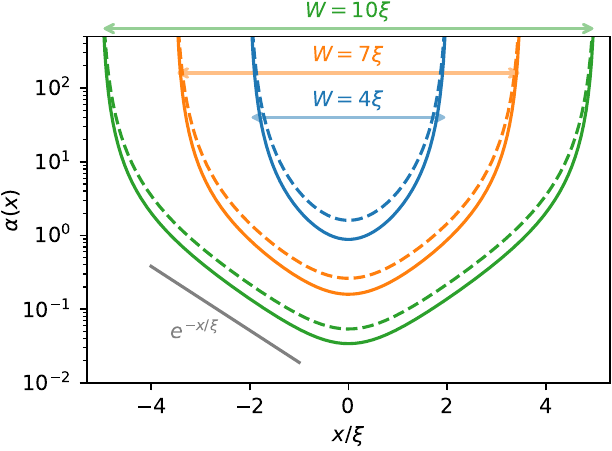}
    \caption{One-dimensional solutions of Eq.~(\ref{eq:main_res}) with boundary values $α(±W/2) = ∞$ for different distances between boundaries $W$. Solid lines show solutions for $η=1$, dashed lines show solutions for $η=0.5$. The gray line shows the exponential decay $\exp(-x/ξ)$. }
    \label{fig:alpha_log}
\end{figure}

To illustrate the solution of Eq.~(\ref{eq:main_res}), we consider the case of an amorphous solid confined by parallel boundaries at the coordinates $x = ±W/2$. Thus, the region $|x| ≤ W/2$ is an unperturbed region with $φ(x) = 0$. The resulting one-dimensional solutions are shown in Fig.~\ref{fig:alpha_x} for different values of the parameter $η$ and different boundary values $α(±W/2)=α_b$. It is notable that there is a solution of Eq.~(\ref{eq:main_res}) with rigid boundaries represented by $α_b=∞$ if $η > 0$. Conversely, there is a solution with the maximally softened limiting boundary value $α_b=-1$ if $η < 1$. Except for these limiting cases, the influence of the parameter $η$ is minor.

The influence of the boundaries decays exponentially with increasing distance from the boundary. This exponential decay is characterized by the length scale $ξ$. To illustrate this behavior, we present the solutions $α(x)$ on a logarithmic scale for $α_b=∞$ and various boundary separations $W$ in Fig.~\ref{fig:alpha_log}.

For each specific geometry of the boundary region, Eq.~(\ref{eq:main_res}) must be solved independently. An illustrative example is provided in Fig.~\ref{fig:halo}. Nevertheless, sufficiently far from the boundaries one has $|α(\Vr)|\ll1$, so that Eq.~(\ref{eq:main_res}) reduces to the linear screened Poisson equation~(\ref{eq:Poisson}). The asymptotic response therefore decays exponentially on the universal length scale $ξ$, with geometry-dependent algebraic prefactors. The solution of the nonlinear Eq.~(\ref{eq:main_res}) for spherical geometry and $η=1$ is presented in Ref.~\cite{Conyuh-effective-elastic-moduli-2023}.

\section{Molecular dynamics simulations}
\label{sec:MD}

\begin{figure}[t!]
    \centering
    \includegraphics[scale=0.8]{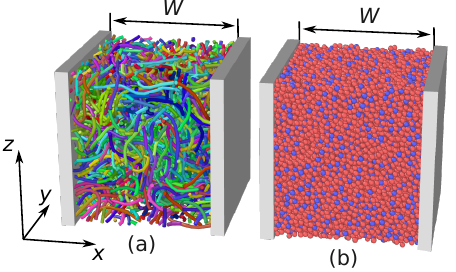}
    \caption{Molecular dynamics configurations of (a) the model polymer and (b) the Lennard-Jones glass confined between rigid boundaries separated by $W=25\,σ$. Only approximately one third of the simulation cell along the $y$ and $z$ directions is shown.}
    \label{fig:MD}
\end{figure}

To test the theoretical predictions for the disorder-induced elastic contrast, we perform athermal quasistatic numerical simulations for two structurally distinct model amorphous systems: a semiflexible bead--spring polymer and a binary Lennard-Jones (LJ) glass (Fig.~\ref{fig:MD}). In both cases, we consider an amorphous layer of thickness $W$ confined between two rigid planes normal to the $x$ axis, with periodic boundary conditions along $y$ and $z$. After averaging over the transverse directions and statistically independent glass realizations, the mechanical response depends only on $x$. This geometry therefore permits a direct comparison with the one-dimensional theoretical profiles
\begin{spalign}
    \label{eq:moduli_x}
    \KK(x) = \KK_0 + \KK_1 α(x),
    \\
    \GG(x) = \GG_0 + \GG_1 α(x),
\end{spalign}
where $α(x)$ is the one-dimensional solution of Eq.~(\ref{eq:main_res}), illustrated in Figs.~\ref{fig:alpha_x} and \ref{fig:alpha_log}.

A spatial variation of the local moduli should be distinguished from an intrinsically nonlocal constitutive response. We therefore determine not only the profiles $\KK(x)$ and $\GG(x)$, but also the corresponding two-point kernels $\KK(x,x')$ and $\GG(x,x')$. This comparison allows us to verify that the microscopic range of elastic nonlocality is much smaller than the heterogeneity length $ξ$ governing the near-boundary stiffening.

\subsection{Model polymer}

We employ the semiflexible Kremer--Grest bead--spring model \cite{
    Kremer-dynamics-entangled-linear-1990,
    Semenov-elastic-properties-nanocomposites-2025}.
Nonbonded beads interact through the repulsive Weeks--Chandler--Andersen potential, whereas consecutive beads are connected by finitely extensible nonlinear elastic bonds. Chain stiffness is introduced through the three-body bending potential
\begin{equation}
    U_{\rm bend}(\theta_{ijk}) = \varkappa(1-\cos\theta_{ijk}),
\end{equation}
where $\theta_{ijk}$ is the angle between two consecutive bonds and $\varkappa$ is the bending-stiffness parameter. This model is widely used to represent semiflexible polymers \cite{
    Egorov-semiflexible-polymers-good-2016,
    Nikoubashman-dynamics-single-semiflexible-2016,
    Sarabadani-driven-translocation-semiflexible-2017}.
All quantities are expressed in the standard LJ units of length $σ$, energy $ϵ$, and time $τ$.

Each system contains $300$ chains of $200$ beads at a monomer number density $ρ=0.85\,σ^{-3}$. The melt is equilibrated following the standard Kremer--Grest protocol, with minor modifications required by the finite bending rigidity $\varkappa=5\,ϵ$. The equilibrated system is cooled to zero temperature at a rate of $8.3\times10^{-4}\,τ^{-1}$ in reduced units, after which the inherent structure is obtained by energy minimization using the FIRE algorithm. A total of $30$ independently cooled and minimized systems were obtained.

\begin{figure}[t!]
    \centering
    \includegraphics[scale=0.8]{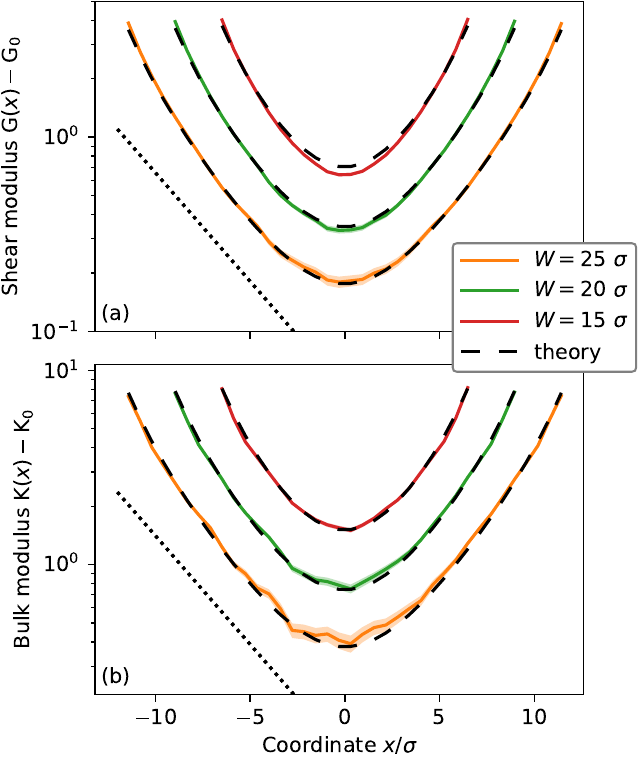}
    \caption{Spatial profiles of (a) the shear modulus $\GG(x)$ and (b) the bulk modulus $\KK(x)$ of the model polymer. Colored bands indicate the estimated statistical uncertainty. Dashed curves show the theoretical profiles given by Eq.~(\ref{eq:moduli_x}) with $α(x)$ obtained from Eq.~(\ref{eq:main_res}) for $η=1$ and $ξ=3.9\,σ$. Dotted lines show the exponential decay $\exp(-x/ξ)$ as a reference. The elastic moduli are given in LJ units $ϵ/σ^3$.}
    \label{fig:KG_local}

    \begin{minipage}[t]{\linewidth}
        \centering
        \makeatletter
        \def\@captype{table}
        \makeatother

        \caption{Parameters obtained by simultaneously fitting the stiffness profiles of the model polymer for three different layer thicknesses $W$.}
        \begin{tabular*}{\columnwidth}{@{\extracolsep{\fill}}@{\hspace{10pt}}cccccc@{\hspace{10pt}}}
            \hline\hline
            $\xi$ & $\KK_0$ & $\GG_0$  & $\KK_1$  & $\GG_1$ & $\alpha_b$\\
            \hline
            $3.9\, \sigma$ & $13.4\, \tfrac{\epsilon}{σ^3}$ & $2.8\, \tfrac{\epsilon}{σ^3}$ & $3.7\, \tfrac{\epsilon}{σ^3}$ & $1.7\, \tfrac{\epsilon}{σ^3}$ & $4.0$ \\[0.5ex]
            \hline\hline
        \end{tabular*}
        \label{tab:kg_fit_results}       
    \end{minipage}
    
\end{figure}

To determine the local elastic moduli, one rigid plane is fixed and a small force corresponding to a prescribed stress $\sigma_{αβ}$ is applied to the other. In the present one-dimensional geometry, the disorder-averaged stress is uniform throughout the mobile layer \cite{
    Semenov-elastic-properties-nanocomposites-2025}.
This is an important simplification relative to the inclusion geometry shown in Fig.~\ref{fig:halo}, for which the disorder-averaged stress remains spatially inhomogeneous.

For normal and shear loading, the plane-averaged displacement field satisfies
\begin{equation}
    \frac{\partial u_x(x)}{\partial x}
    = \frac{\sigma_{xx}}{\KK(x)+\frac{4}{3}\GG(x)},
    \qquad
    \frac{\partial u_y(x)}{\partial x}
    = \frac{\sigma_{xy}}{\GG(x)}.
    \label{eq:local_moduli}
\end{equation}
The displacement gradients are averaged over the $yz$ plane and over independent realizations. The shear response first yields $\GG(x)$, after which $\KK(x)$ is obtained from the normal response. The averaging was done over a total of $5670$ deformation configurations applied to the $30$ cooling seeds by varying the plane positions.

\begin{figure}[t!]
    \centering
    \includegraphics[scale=0.8]{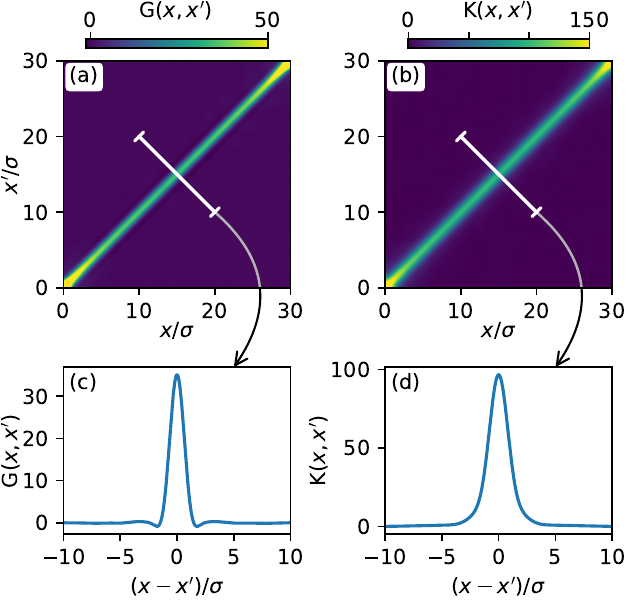}
    \caption{Nonlocal (a) shear elastic kernel $\GG(x,x')$ and (b) bulk elastic kernel $\KK(x,x')$ of the model polymer. Panels (c) and (d) show cross-sections along the white segments in panels (a) and (b), respectively. The nonlocal kernels are given in LJ units $ϵ/σ^4$.}
    \label{fig:KG_nonlocal}
\end{figure}

The resulting profiles are shown in Fig.~\ref{fig:KG_local}. Both moduli increase near the rigid boundaries and approach their bulk values over a distance substantially larger than the interparticle spacing. The enhancement remains measurable even at the center of the widest layer, $12.5\,σ$ from either wall.

We fit all shear- and bulk-modulus profiles simultaneously using Eq.~(\ref{eq:moduli_x}). Over the range relevant to the simulations, the profile shape is only weakly sensitive to $η$ once the amplitudes $\KK_1$ and $\GG_1$ are fitted. Therefore, we set $η=1$, corresponding to a simple form of the elastic contrast given by Eq.~(\ref{eq:alpha1}). The length scale $\xi$, the elastic moduli $\KK_0$, $\GG_0$, $\KK_1$, $\GG_1$, and the boundary value $α_b$ are shared by all curves. Although the confining planes are mechanically rigid, $\alpha_b$ should be regarded as an effective coarse-grained boundary value for the adjacent mobile amorphous material and therefore need not coincide with the idealized limit $\alpha_b\to\infty$. The fitted parameters are given in Table~\ref{tab:kg_fit_results}. The simultaneous fit gives $ξ=3.9\,σ$ and describes all six profiles within their statistical uncertainty. 

We emphasize the deviation of $\KK(x)$ and $\GG(x)$ from the simple exponential decay $\exp(-x/\xi)$, represented by the dotted lines in Fig.~\ref{fig:KG_local}, at large values of $\KK(x)$ and $\GG(x)$. This discrepancy signifies a substantial departure from the behavior predicted by a screened Poisson equation (\ref{eq:Poisson}) in the regime of large $α(x)$. Notably, this deviation is accurately reproduced by the nonlinear equation~(\ref{eq:main_res}), whose solution is used to plot the dashed lines in Fig.~\ref{fig:KG_local}.

The boundary-induced variation of $\KK(x)$ and $\GG(x)$ is conceptually distinct from spatial dispersion, in which the stress at $x$ depends on the strain over a finite neighborhood through the kernels $\KK(x,x')$ and $\GG(x,x')$ \cite{
    Eringen-nonlocal-continuum-field-2004}.
The procedure used to reconstruct these kernels is described in Appendix~\ref{app:nonlocal}. For this calculation, the response was averaged over $3600$ configurations. As shown in Fig.~\ref{fig:KG_nonlocal}, both kernels are strongly concentrated near the diagonal $x=x'$ and become negligible for $|x-x'|\gtrsim2\,σ$. The off-diagonal decay of $\KK(x,x')$ is approximately described by $\exp(-|x-x'|/ξ_{\rm nl})$ at large separations, with $ξ_{\rm nl}\approx 1.1\,σ$. The off-diagonal decay of $\GG(x,x')$ is even faster. The microscopic nonlocality range $ξ_{\rm nl}$ is much shorter than $ξ=3.9\,σ$ and therefore cannot account for the long-range spatial variation of the local moduli.

\subsection{Lennard-Jones glass}

To test whether the predicted behavior depends on polymer connectivity, we also consider the Kob--Andersen binary LJ glass, a standard model of a structurally disordered solid \cite{
    Kob-testing-modecoupling-theory-1995,
    Conyuh-largescale-exponential-correlations-2026}.
The system contains $85^3$ particles, with number fractions $0.8$ and $0.2$ of species A and B, respectively, at a number density $ρ=1.2\,σ^{-3}$. All particles have unit mass and interact through a truncated LJ pair potential with parameters $σ_{AA} = σ$, $σ_{BB} = 0.88\,σ$, $σ_{AB} = 0.8\,σ$, $ϵ_{AA} = ϵ$, $ϵ_{BB} = 0.5\,ϵ$, $ϵ_{AB} = 1.5\,ϵ$, and cutoff distance $2.5\,σ$. Standard LJ units are used throughout.

\begin{figure}[t!]
    \centering
    \includegraphics[scale=0.8]{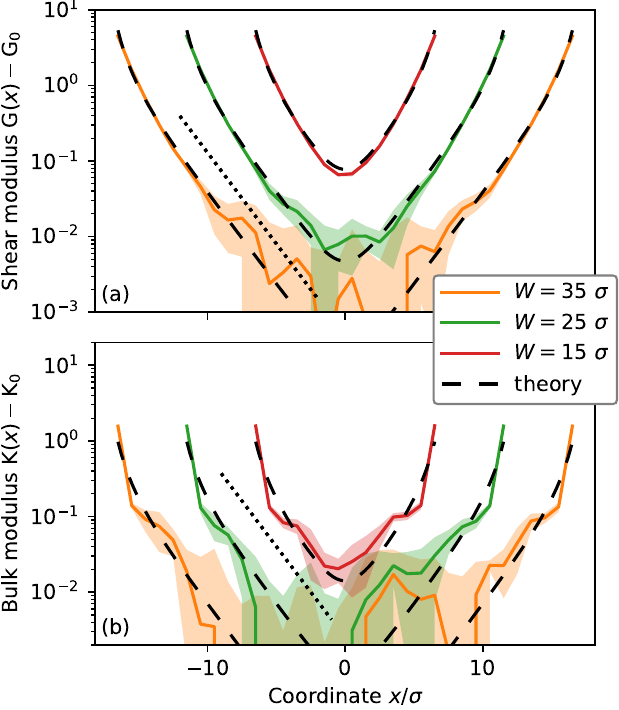}
    \caption{Spatial profiles of (a) the shear modulus $\GG(x)$ and (b) the bulk modulus $\KK(x)$ of the Lennard-Jones glass. Colored bands indicate the estimated statistical uncertainty. Dashed curves show the theoretical profiles given by Eq.~(\ref{eq:moduli_x}) with $α(x)$ obtained from Eq.~(\ref{eq:main_res}) for $η=1$ and $ξ=1.8\,σ$. The dotted line shows the exponential decay $\exp(-x/ξ)$ as a reference. The elastic moduli are given in LJ units $ϵ/σ^3$.}
    \label{fig:LJ_local}
    
    \begin{minipage}[t]{\linewidth}
        \centering
        \makeatletter
        \def\@captype{table}
        \makeatother

        \caption{Parameters obtained by simultaneously fitting the stiffness profiles of the Lennard-Jones glass for three different layer thicknesses.}    
        \begin{tabular*}{\columnwidth}{@{\extracolsep{\fill}}@{\hspace{10pt}}cccccc@{\hspace{10pt}}}
            \hline\hline
            $\xi$ & $\KK_0$ & $\GG_0$  & $\KK_1$  & $\GG_1$ & $\alpha_b$\\
            \hline
            $1.8\, \sigma$ & $64.2\, \tfrac{\epsilon}{σ^3}$ & $19.4\, \tfrac{\epsilon}{σ^3}$ & $0.22\, \tfrac{\epsilon}{σ^3}$ & $1.02\, \tfrac{\epsilon}{σ^3}$ & $287$\\[0.5ex]
            \hline\hline
        \end{tabular*}    
            \label{tab:lj_fit_results}
        \end{minipage}
    \centering
\end{figure}

We use the independently cooled configurations prepared in Ref.~\cite{
    Conyuh-largescale-exponential-correlations-2026}.
Briefly, the initial mixture is heated to the temperature $T=5\,ϵ/k_B$ and cooled in the NVT ensemble at fixed density at a rate of $10^{-3}\,ϵ/k_Bτ$. Cooling is performed in two stages: first to the temperature $T=1\,ϵ/k_B$ with a timestep of $0.005\,τ$, and then to a near-zero temperature with a timestep of $0.01\,τ$. The final zero-temperature equilibrium structures are obtained by FIRE minimization.

The confined geometry and loading protocol are the same as for the polymer. Two groups of particles are constrained to form rigid planes. One plane is fixed, and a small force is applied to the other. After energy minimization, the plane-averaged displacement gradients are converted into $\KK(x)$ and $\GG(x)$ using Eq.~(\ref{eq:local_moduli}). The load is chosen such that the strain in the mobile LJ layer does not exceed $10^{-6}$.

As in Ref.~\cite{
    Conyuh-largescale-exponential-correlations-2026},
configurations dominated by an anomalously soft localized region are excluded from the ensemble average: a configuration is rejected when its largest particle displacement exceeds the mean displacement magnitude by a factor of $100$. This filtering suppresses the disproportionate statistical contribution of rare soft spots and stabilizes the ensemble average. The fitted value of $ξ$ is insensitive to a reasonable change of the cutoff (50, 100, or 200). In total, the resulting averaging was performed over more than $20000$ different configurations, obtained from the $200$ cooling seeds by varying the plane positions.

The spatial profiles presented in Fig.~\ref{fig:LJ_local} again exhibit an enhancement of both elastic moduli in the vicinity of the boundary. The corresponding characteristic decay length is smaller than in the polymeric system. However, within the limits of statistical uncertainty, the profiles are adequately captured by the same theoretical functional form with $ξ ≈ 1.8\,σ$.
The fitted parameters are given in Table~\ref{tab:lj_fit_results}. Again, the length scale $\xi$, the elastic moduli $\KK_0$, $\GG_0$, $\KK_1$, $\GG_1$, and the boundary value $α_b$ are shared by all curves. The apparent deviations at small values of $\GG(x) - \GG_0$ and $\KK(x) - \KK_0$ are attributable to finite sampling statistics in the averaging procedure.

\begin{figure}[t!]
    \centering
    \includegraphics[scale=0.8]{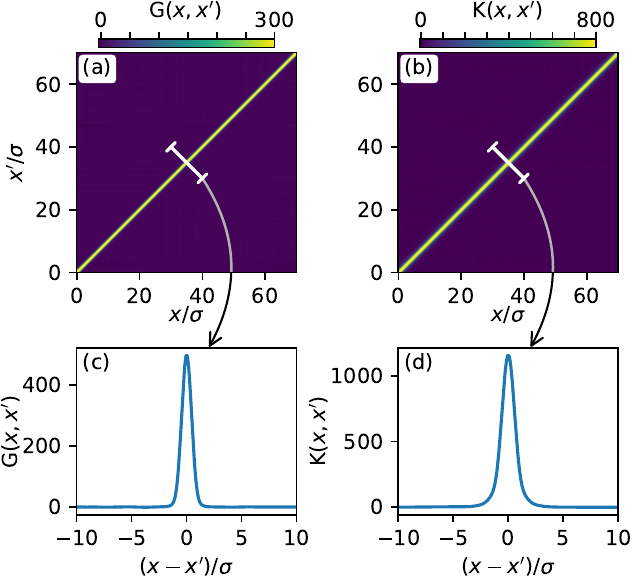}
    \caption{Nonlocal (a) shear elastic kernel $\GG(x,x')$ and (b) bulk elastic kernel $\KK(x,x')$ of the Lennard-Jones glass. Panels (c) and (d) show cross-sections along the white segments in panels (a) and (b), respectively. The nonlocal kernels are given in LJ units $ϵ/σ^4$.}
    \label{fig:LJ_nonlocal}
\end{figure}

Nonlocal kernels for the LJ glass were obtained by averaging over $15300$ configurations and are shown in Fig.~\ref{fig:LJ_nonlocal}. The nonlocal kernels are likewise concentrated near $x=x'$. The off-diagonal decay of $\KK(x,x')$ is approximately described by $\exp(-|x-x'|/ξ_{\rm nl})$ at large separations, with $ξ_{\rm nl}\approx0.55\,σ$. The off-diagonal decay of $\GG(x,x')$ is even faster. Since $ξ_{\rm nl}$ is appreciably smaller than the heterogeneity length scale $ξ ≈ 1.8\,σ$, the LJ glass can also be treated as elastically local on the scale over which the boundary-induced stiffness enhancement develops.

\section{Discussion}
\label{sec:discussion}

The central result of this work is that the spatially resolved, disorder-averaged response of a broad class of stable strongly disordered systems can be described by a scalar contrast field $α(\Vr)$, even though the microscopic degrees of freedom and the corresponding response operators may be vectorial or tensorial. The origin of this simplification is not an assumption of scalar microscopic elasticity. It follows from the spectral structure of the correlation superoperator: in the near-critical strongly disordered regime, the long-range response is dominated by a single nondegenerate upper branch. Near zero wavevector, the dispersion of this branch is isotropic to leading order, while the remaining branches stay separated from the critical value and contribute only on microscopic scales. Consequently, the same slowly varying field $α(\Vr)$ controls the spatial dependence of different components of the effective response. For elasticity, this leads to the common profiles of the bulk and shear moduli in Eq.~(\ref{eq:moduli}), with the microscopic tensorial information retained in the amplitudes $\KK_1$ and $\GG_1$ rather than in separate spatial contrast fields.

For weak perturbations, the upper branch generates the screened Poisson equation~(\ref{eq:Poisson}). Its screening length
$ξ=ξ_0/\sqrt{1-σ_0^2}$ can greatly exceed the structural scale when the largest eigenvalue $σ_0$ approaches unity. Thus, $ξ$ is a collective disorder length rather than a direct measure of pair correlations or the range of the microscopic interactions. The finite-perturbation analysis shows that the same length remains relevant when the contrast is not small. In this regime, the self-consistency of the disorder-averaged Green function produces the nonlinear Eq.~(\ref{eq:main_res}). The nonlinearity of this equation should not be confused with nonlinear material response: the underlying force--displacement relation remains linear, whereas the nonlinear dependence on $α$ originates from averaging and self-consistently inverting a strongly disordered linear-response operator. The parameter $η$ contains microscopic information about the higher-order couplings within the near-critical upper branch. Its influence on the calculated profiles is weak for ordinary finite boundary contrasts, but it determines which limiting solutions, $α_b=-1$ or $α_b\to\infty$, are admissible. Irrespective of $η$, the far-field behavior is universal because Eq.~(\ref{eq:main_res}) reduces to the screened Poisson equation (\ref{eq:Poisson}) for $|α|\ll1$, giving an exponential decay governed by $ξ$.

An important distinction emerges between two different forms of spatial dependence. The effective moduli $\KK(\Vr)$ and $\GG(\Vr)$ vary over the collective length $ξ$, but this does not necessarily imply an intrinsically nonlocal constitutive law over the same distance. In the present description, the response coefficients are local on the scale of interest, while their values depend nonlocally on the surrounding disorder, geometry, and boundary conditions through the solution for $α(\Vr)$. This may be summarized as a local constitutive response with a nonlocal environmental dependence. By contrast, genuine spatial dispersion is described by the two-point kernels $\KK(\Vr,\Vr')$ and $\GG(\Vr,\Vr')$, which directly couple stress and strain at different positions within a microscopic distance $ξ_{\rm nl}\sim ξ_0$. The molecular-dynamics calculations clearly separate these two effects. In the model polymer, the local moduli vary over $ξ=3.9\,σ$, whereas the nonlocal kernels have an off-diagonal decay length
no larger than $ξ_{\rm nl}≈ 1.1\,σ$. In the Lennard-Jones glass, the corresponding scales are $ξ=1.8\,σ$ and $ξ_{\rm nl}\approx 0.55\,σ$. Therefore, in both systems the boundary-induced variation of the local moduli cannot be attributed to the microscopic range of the constitutive kernel.

The simulations also support the interpretation of $ξ$ as an intrinsic property of the disordered matrix. The same theoretical contrast profile describes both the bulk and shear moduli of the confined polymer, using a common value of $ξ$ for different layer thicknesses. The Lennard-Jones glass follows the same functional form but has a smaller fitted length. The numerical values are therefore material dependent, as expected from their relation to the upper eigenvalue branch and to the strength and correlations of disorder. At the same time, the occurrence of the effect in both a linear polymer and a binary atomic glass indicates that it is not tied to a specific polymer architecture or interaction potential. This conclusion is consistent with the observation of large-scale exponential correlations in nonaffine elastic response and with the interpretation of $ξ$ as a nonaffine or heterogeneity length separating microscopic heterogeneous behavior from effective continuum elasticity \cite{
    Conyuh-largescale-exponential-correlations-2026,
    Tanguy-continuum-limit-amorphous-2002,
    Leonforte-continuum-limit-amorphous-2005}.
For the binary Lennard-Jones system, the obtained value $\xi=1.8\sigma$ is of the same order as the nonaffine length of approximately $2.5\sigma$ reported in Ref.~\cite{Conyuh-largescale-exponential-correlations-2026}. The difference is
reasonable given that the latter value was extracted from the tail of the nonaffine correlation function, where the signal-to-noise ratio is relatively low.

Near a rigid inclusion, suppression of nonaffine relaxation produces a smooth, disorder-averaged stiff shell of thickness of order $ξ$. This mechanism provides a microscopic contribution to the interphase commonly introduced in continuum descriptions of nanocomposites \cite{
    Odegard-modeling-mechanical-properties-2005,
    Qiao-simulation-interphase-percolation-2009}.
It is important, however, to distinguish this disorder-induced interphase from a region defined solely by changes in density, chain conformation, or chemical composition. The present mechanism requires no structural perturbation extending throughout the entire stiffened region: a localized constraint can influence nonaffine relaxation collectively over the larger scale $ξ$. This interpretation is compatible with molecular-dynamics results in which the elastic perturbation extends farther from a nanoparticle than the detectable density perturbation \cite{
    Beltukov-local-elastic-properties-2022},
and with other simulations reporting graded mechanical interphases \cite{
    Barakat-predicting-mechanical-heterogeneity-2024}.
In real composites, disorder-induced stiffening will generally coexist with chemical and structural interfacial effects. The present theoretical mechanism isolates the former contribution and predicts how it depends on inclusion geometry, size, and mechanical boundary conditions. In particular, it becomes macroscopically important when the inclusion radius or the spacing between neighboring inclusions is comparable to $ξ$, because the stiffened regions then occupy a substantial fraction of the matrix and may overlap.

The operator formulation developed here is not intrinsically
restricted to elasticity. The same approach can, in principle, be
applied to other stable linear-response problems for which the
disordered response operator admits an analogous positive-semidefinite
random-matrix representation. This opens a route
toward analogous spatial theories for other collective responses of
disordered media, although the physical meaning of the perturbation,
the observable response coefficients, and the appropriate boundary
conditions must be established separately in each case.

A particularly illustrative example is provided by polymer
nanocomposites with an amorphous matrix~\cite{
Cheng-interfacial-properties-polymer-2017}.
The thickness of the interfacial layer, determined from dielectric
spectroscopy, was found to correlate with the characteristic length
scale inferred from the boson peak. The detailed microscopic origin
of this experimentally observed correlation has remained unclear.
The present framework provides a natural explanation for this
observation: the interphase thickness is controlled by the same
disorder length scale $\xi$ that characterizes the collective response
of the amorphous matrix. Within the random-matrix model, both the
boson peak and the Ioffe--Regel crossover emerge at this characteristic
length scale, as discussed in more detail in Refs.~\cite{
Conyuh-largescale-exponential-correlations-2026,
Conyuh-random-matrix-approach-2021,
Beltukov-iofferegel-criterion-diffusion-2013}.

Further developments should also include finite frequencies and
temperatures, anisotropic disorder, compliant or chemically active
interfaces, and direct determination of $\eta$ and $\xi$ from the microscopic correlation superoperator and the properties of its near-critical upper branch rather than from fitted modulus profiles. Such extensions would connect the present static theory to viscoelasticity, wave propagation, and experimentally accessible nanoscale response maps.

\section{Conclusion}
\label{sec:conclusion}

We have developed a general theory for the disorder-averaged nanoscale linear response of stable strongly disordered solids. Mechanical stability is incorporated through a correlated Wishart representation of the response operator, while spatial correlations of disorder are encoded in a correlation superoperator. Its eigenoperator decomposition shows that the long-range response in the near-critical strongly disordered regime is governed by a nondegenerate upper branch. As a result, a generally vectorial or tensorial microscopic problem reduces at long wavelengths to a single scalar disorder-induced contrast $α(\Vr)$.

For a weak perturbation, this contrast satisfies a screened Poisson equation with the collective length $ξ=ξ_0/\sqrt{1-σ_0^2}$. For a finite perturbation, we obtain the nonlinear Eq.~(\ref{eq:main_res}), which retains the same characteristic length and reduces to the previously proposed equation~(\ref{eq:alpha1}) for $η=1$ in an unperturbed region. The nonlinear dependence reflects the self-consistent averaging of a linear disordered system rather than nonlinear elasticity. In the elastic case, the same contrast determines the spatial profiles of both the bulk and shear moduli. The theory therefore provides a modified continuum description in which the effective constitutive coefficients remain local, while their spatial values depend collectively on the surrounding disorder, geometry, and boundary conditions.

Molecular-dynamics simulations of a semiflexible Kremer--Grest polymer and a Lennard-Jones glass support this picture. Rigid boundaries enhance both elastic moduli, and the resulting profiles are described by the theoretical contrast with $ξ=3.9\,σ$ for the polymer and $ξ=1.8\,σ$ for the Lennard-Jones glass. In both systems, the direct nonlocal elastic kernels decay over a shorter microscopic distance. Together with the theory, these results identify the stiffened layer as a collective consequence of suppressed nonaffine relaxation rather than a manifestation of long-range constitutive nonlocality. These results identify $ξ$ as the characteristic thickness of a disorder-induced elastic interphase and provide a bridge between microscopic models of amorphous solids and continuum descriptions of nanostructured disordered materials.

\section{Acknowledgments}

Molecular dynamics simulations were supported by the Russian Science Foundation (grant no. \#22-72-10083-П, \url{https://rscf.ru/en/project/22-72-10083/}). The authors express their gratitude to D.\,A. Conyuh for valuable discussions.

\section{Data availability}
The code that was used for this article is openly available~\cite{Zenodo-nonlinear-response-2026}.

\appendix

\section{Generalized correlated Wishart ensemble}
\label{app:Dyson}

Let us consider the Green function of a general form
\begin{equation}
    \MG = \avgs{(\MA\MV^{-1}\!\MA^\dag + \MW)^{-1}},
\end{equation}
where $\smash{\MA}$ is a zero-mean Gaussian random operator and $\MV$ and $\MW$ are two nonrandom operators. These operators may in general be complex. The deterministic operators $\MV$ and $\MW$ are assumed to be invertible but are not necessarily Hermitian, whereas $\MA$ is, in general, a rectangular operator. In the main text, $\MV$ is the identity operator, whereas in this
appendix we derive the more general formulation, which has a particularly symmetric form. The Green function $\MG$ can be formally expanded in a Neumann series,
\begin{align}
    \MG &= \left<\big(\MA\MV^{-1}\!\MA^\dag + \MW\big)^{-1}\right> 
    \notag\\
    &= \MW^{-1} - \avg{\MW^{-1}\!\MA\MV^{-1}\!\MA^\dag\MW^{-1}} 
    \notag\\
    &\qquad + \avg{ \MW^{-1}\!\MA\MV^{-1}\!\MA^\dag\MW^{-1}\!\MA\MV^{-1}\!\MA^\dag\MW^{-1}} - \ldots
    \label{eq-app:G-series}
\end{align}
Making the interchange $\MA ↔ \MA^\dag$ and $\MW ↔ \MV$, we obtain the secondary Green function
\begin{align}
    \MGb &= \left<\big(\MA^\dag\MW^{-1}\!\MA + \MV\big)^{-1}\right> 
    \notag\\
    &= \MV^{-1} - \avg{\MV^{-1}\MA^\dag\MW^{-1}\!\MA\MV^{-1}}
    \notag\\
    &\qquad + \avg{ \MV^{-1}\!\MA^\dag\MW^{-1}\!\MA\MV^{-1}\!\MA^\dag\MW^{-1}\MA\MV^{-1}} - \ldots
    \label{eq-app:Gb-series}
\end{align}
We use a diagrammatic technique to make the averaging procedure clearer. In particular, we represent the operators as
\begin{equation}
    \begin{aligned}
        \bigl(\MW^{-1}\bigr)_{iα,jβ} &= \newdiag[-10pt]{5}, 
        &
        \bigl(\MA\bigr)_{iα,k} &= \newdiag[-10pt]{7},
        \\
        \bigl(\MV^{-1}\bigr)_{k,l} &= \newdiag[-10pt]{6},
        &
        \bigl(\MA^\dag\bigr)_{k,iα} &= \newdiag[-10pt]{8}.
        \end{aligned}
\end{equation}
The random Gaussian operators $\MA$ and $\MA^\dag$ are represented as open and closed circles, respectively. To perform the averaging, we employ Wick's theorem, which states that all higher-order correlations are represented as a sum of all possible pairwise covariances. Each pair is represented as a dashed line connecting the two points:
\begin{align}
    \mathcal{C}_{iα,jβ;k,l} &= \avg{\bigl(\MA\bigr)_{iα,k} \bigl(\MA^\dag\bigr)_{l,jβ}} = \newdiag[-10pt]{9},
    \\
    \mathcal{P}_{iα,l;k,jβ} &= \avg{\bigl(\MA\bigr)_{iα,k} \bigl(\MA\bigr)_{jβ,l}} = \newdiag[-10pt]{10},
    \\
    (\mathcal{P}^\dag)_{k,jβ;iα,l} &= \avg{\bigl(\MA^\dag\bigr)_{k,iα} \bigl(\MA^\dag\bigr)_{l,jβ}} = \newdiag[-10pt]{11}.
\end{align}
The first one is the normal covariance. The other two are pseudocovariances (anomalous covariances). For conventional time-reversal-symmetric elastic systems, $\MA$ may be chosen real. In this case, the pseudocovariances coincide with the normal covariance up to the corresponding index rearrangement.

Therefore, a graphical representation of the Green functions $\MG$ and $\MGb$ is
\begin{align}
    \newdiag{1} &= 
    \newdiag{3} - \newdiag{12} + \newdiag{14} 
    \notag\\
    &\quad + \newdiag{16} + \newdiag{18} - \dots    
    \label{eq-app:G-diagram}
    \\
    \newdiag{2} &= 
    \newdiag{4} - \newdiag{13} + \newdiag{15} 
    \notag\\
    &\quad + \newdiag{17} + \newdiag{19} - \dots    
    \label{eq-app:Gb-diagram}
\end{align}
where the order of operators is the same as in
Eqs.~(\ref{eq-app:G-series}) and (\ref{eq-app:Gb-series}).
The presentation in Eqs.~(\ref{eq-app:G-diagram}) and
(\ref{eq-app:Gb-diagram}) allows us to distinguish planar and
nonplanar contractions. Since $\MA$ and $\MA^\dag$ alternate in these
series, contractions involving the pseudocovariances $\mathcal P$ and
$\mathcal P^\dag$ necessarily generate nonplanar diagrams, whereas the
planar diagrams involve only the normal covariance $\mathcal C$.

We use the same planar approximation as in
Ref.~\cite{Conyuh-largescale-exponential-correlations-2026}.
As discussed there, each closed index loop produces a trace factor
$T_Λ$. For a fixed number of contractions, planar diagrams
contain the maximal number of closed loops, while nonplanar diagrams
contain fewer loops. Therefore, in the regime $T_Λ\gg1$, planar
diagrams give parametrically larger contributions than nonplanar
diagrams of the same order. This condition does not require $\MA$ to
be dense and may also hold for sparse matrices when each generalized
bond involves a sufficiently large number of degrees of freedom.
Retaining only the leading planar diagrams, we obtain the two Dyson
series:
\begin{align}
    \newdiag{1} &= 
    \newdiag{3} - \newdiag{20} + \newdiag{22}  - \dots    
    \\
    \newdiag{2} &= 
    \newdiag{4} - \newdiag{21} + \newdiag{23}  - \dots    
\end{align}
Summing these series gives
\begin{equation}
    \MG = \bigl(\FMC\,\MGb + \MW\bigr)^{-1}, 
    \quad  
    \MGb = \bigl(\FMC^\dag \MG + \MV\bigr)^{-1},
    \label{eq-app:G-closed2}
\end{equation}
where the action of the correlation superoperators $\FMC$ and $\FMC^\dag$ on the Green functions $\MGb$ and $\MG$, respectively, has the following graphical representation
\begin{align}
    \FMC &= \newdiag[-14pt]{38}, 
    \quad
    \FMC\,\MGb = \newdiag[-14pt]{24},
    \\
    \FMC^\dag &= \newdiag[-14pt]{40},
    \quad
    \FMC^\dag \MG = \newdiag[-14pt]{25}.
\end{align}
Although $\FMC$ and $\FMC^\dag$ have the same graphical representation
(black and white dots connected by a dashed line), their input and
output legs belong to opposite operator spaces. We therefore draw the
superoperators vertically, with the input legs on the right.

Finally, taking $\MV = \M1$, we obtain the Dyson equations used in the main text in Eq.~(\ref{eq:Dyson}).

\section{Eigenoperator decomposition of the correlation superoperator}
\label{app:eigval}

\subsection{Factorization}

For $\omega=i\zeta$ with $\zeta>0$, the Green functions $\MG$
and $\MGb$ are Hermitian and positive definite. Therefore, the
corresponding propagation superoperators $\FMR$ and $\FMRb$ are
also Hermitian and positive definite. The static limit is taken
subsequently as $\zeta\to0^+$. Indeed, the corresponding quadratic forms are positive for arbitrary nonzero operators $\MX$ and $\MXb$:
\begin{spalign}
    \Tr{\MX^\dag\FMR\MX} &= \Tr{\MX^\dag\MG\MX\MG} = \Tr{\MY^\dag\MY} > 0,
    \\
    \Tr{\MXb^\dag\FMRb\MXb} &= \Tr{\MXb^\dag\MGb\MXb\MGb} = \Tr{\MYb^\dag\MYb} > 0,
\end{spalign}
where we have introduced
\begin{equation}
    \MY = \MG^{1/2}\MX\thin\MG^{1/2}, \qquad \MYb = \MGb^{1/2}\MXb\thin\MGb^{1/2}.
\end{equation}
It can be readily verified that the superoperator
\begin{equation}
    \FM B = \FMC\,\FMRb\,\FMC^\dag = \newdiag[-14pt]{28}
\end{equation}
is Hermitian and positive semidefinite. Since $\FMR$ is Hermitian and
positive definite, the product $\FM B\FMR$ is diagonalizable and has
real nonnegative eigenvalues~\cite[Corollary~7.6.2]{Horn-matrix-analysis-2017}.
We therefore introduce eigenoperators $\MS_v$ satisfying
\begin{equation}
    \label{eq:B-eigval-def}
    \FM B\,\FMR\,\MS_v = θ_v\MS_v,
\end{equation}
with $θ_v\geq0$. The eigenoperators $\MS_v$ play the same role in the
operator space as eigenvectors do in an ordinary vector space. The graphical representation of Eq.~(\ref{eq:B-eigval-def}) is
\begin{equation}
    \newdiag[-14pt]{29} = \newdiag[-14pt]{30},
    \label{eq:B-eigval}
\end{equation}
where the blue open triangle denotes the operator $\MS_v$ and the thick black wavy line denotes the eigenvalue $θ_v$. These eigenoperators satisfy the $\FMR$-weighted orthogonality condition
\begin{equation}
    \label{eq:orthog_R}
    \Tr{\MS^\dag_v\,\FMR\,\MS_{v'}} = \newdiag[-14pt]{31} = δ_{vv'},
\end{equation}
where the blue filled triangle represents $\MS^\dag_v$.

For each nonzero eigenvalue $θ_v$, we define
\begin{equation}
    σ_v = \sqrt{θ_v},
    \qquad
    \MSb_v = σ_v^{-1}\FMC^\dag\FMR\,\MS_v.
\end{equation}
These definitions yield the symmetrized eigenvalue problem
\begin{spalign}
    \label{eq:eigval}
    \FMC\,\FMRb\thin\MSb_v &= σ_v \MS_v,
    \\
    \FMC^\dag\FMR\,\MS_v &= σ_v \MSb_v.
\end{spalign}
The pair of eigenoperators $\MS_v$ and $\MSb_v$ defines a disorder-mediated response channel parameterized by $σ_v=\sqrt{θ_v} > 0$. For simplicity, throughout the paper we refer to $σ_v$ as the eigenvalues of the symmetrized problem~(\ref{eq:eigval}), although in standard linear-algebra terminology they are the nonzero singular values of the propagation-weighted correlation superoperator $\FMR^{1/2}\FMC\FMRb^{1/2}$. Operators $\MS_v$ and $\MSb_v$ span only the disorder-mediated subspace of $N_\dof × N_\dof$ and $N_\bb × N_\bb$ operator spaces, respectively.

The graphical representation of equations (\ref{eq:eigval}) is
\begin{spalign}
    \newdiag[-14pt]{35} &= \newdiag[-14pt]{36},
    \\
    \newdiag[-14pt]{33} &= \newdiag[-14pt]{34},
\end{spalign}
where the open red triangles denote the operator $\MSb_v$ and the thin black wavy line denotes the eigenvalue $σ_v$.

The second orthogonality relation is
\begin{multline}
    \label{eq:orthog_Rb}
    \Tr{\MSb^\dag_v\,\FMRb\,\MSb_{v'}} = \newdiag[-14pt]{37}
    \\
    = σ_v^{-1}σ_{v'}^{-1}\Tr{\MS^\dag_v\thin\FMR\,\FMC\,\FMRb\thin\FMC^\dag\FMR\,\MS_{v'}} = δ_{vv'},
\end{multline}
where the red filled triangle represents $\MSb^\dag_v$. Thus, the operators $\MS_v$ and $\MSb_v$ form propagation-weighted orthonormal bases of the corresponding disorder-mediated subspaces.

Finally, we can factorize the correlation superoperator $\FMC$ and its adjoint $\FMC^\dag$ as
\begin{spalign}
    \label{eq:C_fact}
    \FMC = \newdiag[-14pt]{38} &= \sum_v \newdiag[-14pt]{39},
    \\
    \FMC^\dag = \newdiag[-14pt]{40} &= \sum_v \newdiag[-14pt]{41}.
\end{spalign}
This presentation means that we can separate the influence of disorder into different disorder-mediated response channels enumerated by $v$.


Using the factorization (\ref{eq:C_fact}) with the orthogonality condition (\ref{eq:orthog_Rb}), the superoperator $\FM B$ can also be factorized:
\begin{multline}
    \FM B = \FMC\,\FMRb\,\FMC^\dag = \newdiag[-14pt]{28}
    \\
    = \sum_{vv'} \newdiag[-14pt]{45} = \sum_v \newdiag[-14pt]{42}.
\end{multline}
As in Eq.~(\ref{eq:B-eigval}), the thick wavy line denotes the squared eigenvalue $σ_v^2 = θ_v$.

The effective force-constant matrix given by Eq.~(\ref{eq:Phi_eff_series}) can be written using the superoperator $\FM B$ as
\begin{equation}
    δ\M{Φ}^\eff = δ\M{Φ} + \FM B \FMR\,δ\M{Φ} + \FM B \FMR\,\FM B \FMR\,δ\M{Φ} + \ldots.
\end{equation}
This series has a familiar ladder form, presented in Fig.~\ref{fig:ladder}. This type of diagram is known as a \emph{diffuson} \cite{
    Rammer-quantum-transport-theory-2018,
    Sadovskii-diagrammatics-lectures-selected-2006}.
It corresponds to ``twisted'' diagrams in our previous work \cite{
    Conyuh-largescale-exponential-correlations-2026}
where the parallel orientation of propagators was assumed.

Summing this geometric series, we obtain
\begin{equation}
    δ\M{Φ}^\eff = δ\M{Φ} + \sum_v \newdiag[-14pt]{46},
\end{equation}
where the double wavy line is a ``dressed'' scattering kernel
\begin{equation}
    D_v = \frac{σ_v^2}{1-σ_v^2},
\end{equation}
which represents diffusion through the disorder. This kernel is nonnegative because $σ_v\leq1$ (see Appendix~\ref{app:herm_min}).

\subsection{Bloch's theorem}
\label{app:Bloch}

For a statistically homogeneous amorphous medium without boundaries,
disorder averaging restores translational invariance on the reference
lattice.

Let us consider the translation operator $\MT_{\Vr}$, where ${\Vr}$ denotes a lattice vector connecting sites of the reference lattice. This operator may act both in the space of degrees of freedom and in the secondary bond space. For the statistically homogeneous system under consideration, the Green functions commute with the translation operator
\begin{spalign}
    \MT_{\Vr}\thin\MG &= \MG\,\MT_{\Vr},
    \\
    \MT_{\Vr}\thin\MGb &= \MGb\thin\MT_{\Vr}.
\end{spalign}

The associated superoperator is defined as
\begin{equation}
    \FMT_{\Vr}\thin \MX = \MT_{\Vr} \thin \MX\,\MT_{\Vr}^\dag
\end{equation}
for any ordinary operator $\MX$ both in the space of degrees of freedom and in the secondary bond space. Since $\MT_{\Vr}^\dag = \MT_{-\Vr}$, the same applies to the superoperator
\begin{equation}
    \FMT_{\Vr}^{\thin\dag} = \FMT_{-\Vr}.
\end{equation}
The translation superoperator commutes with the propagation superoperators,
\begin{spalign}
    \FMT_{\Vr}\,\FMR &= \FMR\,\FMT_{\Vr},
    \\
    \FMT_{\Vr}\,\FMRb &= \FMRb\FMT_{\Vr}
\end{spalign}
and the correlation superoperator,
\begin{equation}
    \FMT_{\Vr}\,\FMC = \FMC\,\FMT_{\Vr}.
\end{equation}

Thus, the eigenoperators of Eq.~(\ref{eq:eigval}) can be chosen as eigenoperators of the translation superoperators
\begin{spalign}
    \FMT_{\Vr}\MS_v &= τ_{v\Vr} \MS_v,
    \\
    \FMT_{\Vr}\MSb_v &= τ_{v\Vr} \MSb_v
    \label{eq:eigval_tr}
\end{spalign}
with some eigenvalue $τ_{v\Vr}$. According to Bloch's theorem~\cite{Ashcroft-solid-state-physics-1976, Esposito-emergence-diffusion-finite-2005}, for each eigenvalue index $v$ there exists a wavevector $\Vp$ such that
\begin{equation}
    τ_{v\Vr} = e^{-i\Vp\Vr}.
\end{equation}
Consequently, each eigenvalue index $v$ can be written as a pair $(n,\Vp)$, where $n$ denotes the branch index and $\Vp$ the Bloch wavevector.

\subsection{Hermitian conjugation}
\label{app:Herm}

To further elaborate on the properties of eigenoperators $\MS_{n\Vp}$ and $\MSb_{n\Vp}$, we consider the antilinear Hermitian-conjugation superoperator
\begin{equation}
    \FMK\MX = \MX^\dag,
\end{equation}
where $\MX$ is any ordinary operator (in the direct $N_\dof×N_\dof$ space or in the auxiliary $N_\bb×N_\bb$ space depending on the context). The conjugation superoperator $\FMK$ commutes with the propagation superoperators $\FMR$ and $\FMRb$
\begin{gather}
    \FMK\,\FMR\,\MX = (\MG\MX\MG)^\dag = \MG\MX^\dag\MG = \FMR\,\FMK\,\MX,
    \\
    \FMK\,\FMRb\thin\MX = (\MGb\MX\MGb)^\dag = \MGb\MX^\dag\MGb = \FMRb\thin\FMK\,\MX
\end{gather}
and the correlation superoperator
\begin{equation}
    \FMK\,\FMC\thin\MX = \FMK\,\avgs{\MA\MX\MA^\dag} = \avgs{\MA\MX^\dag\MA^\dag} = \FMC\,\FMK\thin\MX.
\end{equation}
However, the conjugation superoperator does not commute with the translation operator in the general case
\begin{equation}
    \FMK\,\FMT_{\Vr} = \FMT_{-\Vr}\,\FMK
\end{equation}
since the complex conjugation reverses the direction of translation. Hermitian conjugation therefore maps an eigenoperator with wavevector $\Vp$ onto an eigenoperator with wavevector $-\Vp$ and the same eigenvalue. By an appropriate choice of the phases of the eigenoperators and, within degenerate subspaces, a suitable choice of basis, we may impose the convention
\begin{spalign}
    \FMK\,\MS_{n\Vp} ≡ \MS^\dag_{n\Vp} &= \MS_{n,-\Vp},
    \\
    \FMK\,\MSb_{n\Vp} ≡ \MSb^\dag_{n\Vp} &= \MSb_{n,-\Vp}.
\end{spalign}
This convention requires the use of a complex basis, even if the initial problem is real.

\section{Dyson equations as a minimization}
\label{app:minimum}

\subsection{General case}

It is well known that Dyson equations can be obtained within a saddle-point approximation to an effective action~\cite{John-localization-disordered-elastic-1983, Belitz-andersonmott-transition-1994}. In our case, the following functional of the operators $\Mg$ and $\Mgb$ can be introduced:
\begin{multline}
    ℱ(\Mg, \Mgb) = -\ln\det\Mg - \ln\det\Mgb
    \\
    + \Tr{\Mg\,\FMC\thin\Mgb} + \Tr{\Mg\thin\MW} + \Tr\Mgb.  \label{eq:F}
\end{multline}
In general, it is a complex holomorphic function of two operators $\Mg$ and $\Mgb$, which do not necessarily have to be Hermitian. We now demonstrate that a saddle point of this functional coincides with the solution of the Dyson equations given in Eq.~(\ref{eq:Dyson}). To this end, we consider the expansion of $ℱ$ in a power series:
\begin{equation}
    ℱ\bigl(\MG + δ\Mg, \MGb + δ\Mgb\bigr) = ℱ^{(0)} + ℱ^{(1)} + ℱ^{(2)} + \ldots,  \label{eq:F_series}
\end{equation}
in the vicinity of the saddle point $(\Mg,\Mgb)=(\MG,\MGb)$, where $ℱ^{(0)} = ℱ(\MG, \MGb)$ denotes the saddle-point value. The first-order term is~\cite{Hjorungnes-complexvalued-matrix-derivatives-2011}
\begin{multline}
    ℱ^{(1)} = \Tr{\bigl(-\MG^{-1} + \FMC\,\MGb + \MW\bigr)δ\Mg}  + 
    \\
    \Tr{\bigl(-\MGb^{-1} + \FMC^\dag\MG + \M1\bigr) δ\Mgb},
\end{multline}
At the saddle point, the condition $ℱ^{(1)} = 0$ must hold for arbitrary independent variations $δ\Mg$ and $δ\Mgb$. This stationarity condition is equivalent to the Dyson equations (\ref{eq:Dyson}).

The second-order term $ℱ^{(2)}$ is
\begin{multline}
    \label{eq:F2}
    ℱ^{(2)} = \frac{1}{2}\Tr{\MG^{-1}δ\Mg\,\MG^{-1}δ\Mg}
     + \frac{1}{2}\Tr{\MGb^{-1}δ\Mgb\,\MGb^{-1}δ\Mgb}
    \\    
    + \Tr{δ\Mg\,\FMC\,δ\Mgb}.
\end{multline}
Higher-order terms $ℱ^{(k)}$ with $k≥3$ are
\begin{equation}
    ℱ^{(k)} = \frac{(-1)^k}{k}\Big(\Tr{\big(\MG^{-1}δ\Mg\big)^k}
    + \Tr{\big(\MGb^{-1}δ\Mgb\big)^k}\Big).
\end{equation}

\subsection{Minimization over Hermitian matrices}
\label{app:herm_min}

In what follows, we consider the static response regularized by
$\omega=i\zeta$ with $\zeta>0$. In this case,
$\MW=\M{Φ}_0+\zeta^2\M m$ is Hermitian and positive definite,
and the physical solutions $\MG$ and $\MGb$ of the Dyson equations are
Hermitian and positive definite.

We now examine the variational character of this solution by
considering $ℱ(\Mg,\Mgb)$ on the domain of Hermitian
positive-definite operators $\Mg$ and $\Mgb$. On this domain,
$ℱ(\Mg,\Mgb)$ is real-valued. We show that the physical saddle
point $\Mg=\MG$, $\Mgb=\MGb$ is the minimum of this functional.

It is crucial that $ℱ(\Mg,\Mgb)$ possesses a unique minimum
within the cone of positive-definite operators $\Mg$ and $\Mgb$. If
this were not the case, there would exist either multiple or no
positive-definite solutions $\MG$ and $\MGb$ to the Dyson
equations~(\ref{eq:Dyson}), in contradiction with the results of
Ref.~\cite{Helton-operatorvalued-semicircular-elements-2007}.

At the minimum $\Mg=\MG$ and $\Mgb=\MGb$, the second-order variation
$ℱ^{(2)}$ must be nonnegative for arbitrary Hermitian
variations. To establish an upper bound on the eigenvalues
$σ_{n\Vp}$, it is sufficient to consider variations restricted to the
disorder-mediated subspaces spanned by the eigenoperators
$\MS_{n\Vp}$ and $\MSb_{n\Vp}$. Using the orthogonality
relations~(\ref{eq:orthog}) and the factorization of the correlation
superoperator, Eq.~(\ref{eq:F2}) gives
\begin{multline}
    ℱ^{(2)}
    = \frac{1}{2}\sum_{n\Vp}
    \big(
        δa_{n\Vp}\,δa_{n,-\Vp}
        + δb_{n\Vp}\,δb_{n,-\Vp}
    \\
        + 2σ_{n\Vp}\,δa_{n\Vp}\,δb_{n,-\Vp}
    \big),
\end{multline}
where the expansion coefficients are defined by
\begin{spalignat}{2}
    δ\Mg &= \FMR \sum_{n\Vp} δa_{n\Vp}\MS_{n\Vp},
    &
    \quad
    δa_{n\Vp} &= \Tr{\MS_{n\Vp}^\dag\thinδ\Mg},
    \\
    δ\Mgb &= \FMRb \sum_{n\Vp} δb_{n\Vp}\MSb_{n\Vp},
    &
    \quad
    δb_{n\Vp} &= \Tr{\MSb_{n\Vp}^\dag\thinδ\Mgb}.
\end{spalignat}
For Hermitian variations, Eq.~(\ref{eq:Snp_Herm}) implies
$δa_{n,-\Vp}=δa_{n\Vp}^*$ and
$δb_{n,-\Vp}=δb_{n\Vp}^*$. Consequently, the quadratic form
associated with each disorder-mediated response channel is
nonnegative only if $1-σ_{n\Vp}^2 \geq 0$. Since $σ_{n\Vp}>0$ by construction, we obtain $0 < σ_{n\Vp} \leq 1$.

\subsection{Single argument (a)}

The functional $ℱ(\Mg, \Mgb)$ is convex with respect to a positive-definite Hermitian operator $\Mgb$ for fixed $\Mg$. It possesses a unique minimum at
\begin{equation}
    \Mgb = \operatorname{argmin}_{\Mgb} ℱ(\Mg, \Mgb) = \big(\FMC^\dag\Mg + \M1\big)^{-1},
\end{equation}
which coincides exactly with the second Dyson equation~(\ref{eq:Dyson}). Substituting this minimizing value of $\Mgb$ into Eq.~(\ref{eq:F}), we obtain
\begin{multline}
    ℱ_\aa(\Mg) ≡ \min_{\Mgb} ℱ(\Mg, \Mgb)
    \\
    = -\ln \det \Mg + \ln \det (\FMC^\dag\Mg + \M1) + \Tr{\Mg\MW} + N_\bb.
\end{multline}
We consider the power-series expansion
\begin{equation}
    ℱ_\aa(\MG + δ\Mg) = ℱ_\aa^{(0)} + ℱ_\aa^{(1)} + ℱ_\aa^{(2)} + \ldots  \label{eq:Fa_series}
\end{equation}
in a neighborhood of the optimal point $\Mg = \MG$, where $ℱ_{\smash{\aa}}^{(0)} = ℱ_\aa(\MG)$ denotes the corresponding optimal value. At the minimum, the first-order contribution $ℱ_{\smash{\aa}}^{(1)}$ vanishes. The higher-order terms of the expansion are given by
\begin{equation}
    ℱ_\aa^{(k)} = \frac{(-1)^k}{k}\Bigl(\Tr{\big(\MG^{-1}δ\Mg\big)^k}
    - \Tr{\big(\MGb\,\FMC^\dagδ\Mg\big)^k}\Bigr)
\end{equation}
for $k ≥ 2$.

\subsection{Single argument (b)}

The function $ℱ(\Mg, \Mgb)$ is a convex function of a positive-definite Hermitian operator $\Mg$ for a fixed $\Mgb$. It has a single minimum, given by
\begin{equation}
    \Mg = \operatorname{argmin}_{\Mg} ℱ(\Mg, \Mgb) = \big(\FMC\,\Mgb + \MW\big)^{-1},
\end{equation}
which is exactly the first Dyson equation (\ref{eq:Dyson}). 
Substituting this minimizing value of $\Mg$ into Eq.~(\ref{eq:F}), we obtain
\begin{multline}
    ℱ_\bb(\Mgb) ≡ \min_{\Mg} ℱ(\Mg, \Mgb)
    \\
    = - \ln \det \Mgb + \ln \det (\FMC\Mgb + \MW) + \Tr{\Mgb} + N_\dof.
\end{multline}
We consider the power-series expansion
\begin{equation}
    ℱ_\bb(\MGb + δ\Mgb) = ℱ_\bb^{(0)} + ℱ_\bb^{(1)} + ℱ_\bb^{(2)} + \ldots   \label{eq:Fb_series}
\end{equation}
in a neighborhood of the optimal point $\Mgb = \MGb$, where $ℱ_\bb^{(0)} = ℱ_\bb(\MGb)$ denotes the corresponding optimal value. At the minimum, the first-order contribution $ℱ_\bb^{(1)}$ vanishes. The higher-order terms of the expansion are given by
\begin{equation}
    ℱ_\bb^{(k)} = \frac{(-1)^k}{k}\Bigl(\Tr{\big(\MGb^{-1}δ\Mgb\big)^k}
    - \Tr{\big(\MG\,\FMC\,δ\Mgb\big)^k}\Bigr)
\end{equation}
for $k ≥ 2$.

\subsection{Iterative minimization}
While the function $ℱ(\Mg, \Mgb)$ is not a convex function of both arguments simultaneously, we can use the following iteration procedure
\begin{spalign}
    \label{eq:iteration}
    \Mg^{(n+1)} &= \left(\FMC\,\Mgb^{(n)} + \MW\right)^{-1}, 
    \\
    \Mgb^{(n+1)} &= \left(\FMC^\dag\Mg^{(n+1)} + \M1\right)^{-1}
\end{spalign}
to find a solution to the Dyson equations~(\ref{eq:Dyson}) using any positive-definite operator $\Mgb^{(0)}$ as a starting point. This iterative scheme generates the monotone sequence $ℱ\bigl(\Mg^{(n)}, \Mgb^{(n)}\bigr) ≥ ℱ\bigl(\Mg^{(n+1)}, \Mgb^{(n)}\bigr) ≥ ℱ\bigl(\Mg^{(n+1)}, \Mgb^{(n+1)}\bigr)$. Equality holds only when the fixed point is reached. Since it is known that the Dyson equation admits a unique solution for any $σ_0<1$, fixed-point iteration (\ref{eq:iteration}) converges to that unique solution.

\section{Moments of the near-critical channel}

Let us consider the moments of the zero-wavevector channel ($n = 0$, $\Vp = 0$) of the upper branch
\begin{spalign}
    P_k &= \Tr{\big(\MG\,\MS_{00}\big)^k},
    \\
    Q_k & = \Tr{\big(\MGb\,\MSb_{00}\big)^k}
\end{spalign}
for integers $k\ge1$. Normalization (\ref{eq:orthog}) of the eigenoperators $\MS_{00}$ and $\MSb_{00}$ gives
\begin{equation}
    \label{eq:PQ2}
    P_2 = 1, \qquad Q_2 = 1.
\end{equation}
Third moments are related to the nonlinear coefficients introduced in Eq.~(\ref{eq:PQ}) as
\begin{equation}
    P_3 = P_{000}^{000}, \qquad Q_3 = Q_{000}^{000}.
\end{equation}
We now ask what values the moments $P_k$ and $Q_k$ can take, especially in the near-critical regime $σ_0\to1$.

\subsection{Lower bounds}

The upper eigenoperators $\MS_{00}$ and $\MSb_{00}$ can be chosen
positive semidefinite by the Perron--Frobenius theorem for the
corresponding positive superoperators. Since the operators $\MG$ and $\MGb$ are positive definite, we may consider the positive semidefinite operators
\begin{equation}
    \MG^{1/2}\,\MS_{00}\,\MG^{1/2},
    \qquad
    \MGb^{1/2}\,\MSb_{00}\,\MGb^{1/2},
\end{equation}
and denote their eigenvalues by $λ_i$ and $μ_i$, respectively. Therefore, 
\begin{equation}
    P_k = \sum_i λ_i^k,
    \qquad
    Q_k = \sum_i μ_i^k.
\end{equation}
Using the Cauchy–Schwarz inequality and nonnegativity of $λ_i$, we obtain
\begin{multline}
    \label{eq:Pn_ineq}
    P_k^2 = \Big(\sum_i λ_i^{(k-1)/2}λ_i^{(k+1)/2}\Big)^2 
    \\
    ≤ \sum_i λ_i^{k-1} \sum_i λ_i^{k+1} = P_{k-1}P_{k+1}.
\end{multline}
Thus, the moments $P_k$ form a log-convex sequence. Applying the same reasoning to the eigenvalues $μ_i$ yields an analogous inequality for $Q_k$:
\begin{equation}
    Q_k^2 ≤ Q_{k-1}Q_{k+1}.
\end{equation}

Combining these inequalities with Eq.~(\ref{eq:PQ2}), we arrive at the bounds
\begin{equation}
    P_k ≥ P_1^{2-k}, \qquad Q_k ≥ Q_1^{2-k}.
    \label{eq:PQ_lower_bound}
\end{equation}

\subsection{Ratio}

To elaborate on the ratio between the moments $P_k$ and $Q_k$, we consider the minimization scheme given in Appendix~\ref{app:minimum}. The functional $ℱ(\MG + δ\Mg, \MGb + δ\Mgb)$ should not be smaller than $ℱ_0$ for any small $δ\Mg$ and $δ\Mgb$.

Let us consider
\begin{equation}
    δ\Mg = δa\,\FMR\,\MS_{00}   \label{eq:dg}
\end{equation}
where $δa$ is a small real coefficient and $δ\Mgb$ is chosen to minimize the functional for the given $δ\Mg$. In this case $ℱ(\MG + δ\Mg, \MGb + δ\Mgb) = ℱ_\aa(\MG + δ\Mg) = ℱ_\aa^{(0)} + δℱ_\aa$. Taking into account that $\FMC^\dagδ\Mg = σ_0\,\MSb_{00}\,δa$ by Eq.~(\ref{eq:S_eigval}), we obtain
\begin{equation}
    δℱ_\aa = \sum_{k=2}^{∞}\frac{(-1)^k}{k}δa^k \Big( P_k - σ_0^k Q_k\Big).
\end{equation}

Let us consider
\begin{equation}
    δ\Mgb = δb\,\FMRb\,\MSb_{00}  \label{eq:dgb}
\end{equation}
where $δb$ is a small real coefficient and $δ\Mg$ is chosen to minimize the functional for the given $δ\Mgb$. In this case $ℱ(\MG + δ\Mg, \MGb + δ\Mgb) = ℱ_\bb(\MGb + δ\Mgb) = ℱ_\bb^{(0)} + δℱ_\bb$. Taking into account that $\FMCδ\Mgb = σ_0\,\MS_{00}\,δb$ by Eq.~(\ref{eq:S_eigval}), we obtain
\begin{equation}
    δℱ_\bb = \sum_{k=2}^{∞}\frac{(-1)^k}{k}δb^k \Big( Q_k - σ_0^k P_k\Big).
\end{equation}

Both $δℱ_\aa$ and $δℱ_\bb$ must be nonnegative for arbitrary values of $δa$ and $δb$. It follows that $δℱ_\aa \to -\,δℱ_\bb$ in the limit $σ_0 \to 1$ under the condition $δb = δa$. Therefore, in the critical system characterized by $σ_0 = 1$, we obtain $δℱ_\aa = δℱ_\bb = 0$ and thus
\begin{equation}
    P_k = Q_k, \quad k ≥ 2.
    \label{eq:PQ_eq}
\end{equation}



\subsection{Critical values}

\begin{figure}
    \centering
    \includegraphics[scale=0.8]{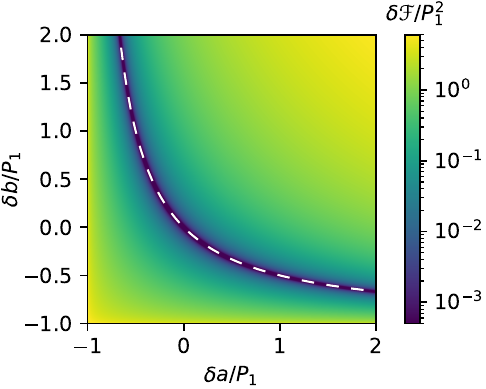}
    \caption{
    Increase in the functional $δℱ$ as a function of the coefficients $δa$ and $δb$ for $σ_0=1$, according to Eq.~(\ref{eq:delta_F_crit}). The white dashed line represents the critical line $δℱ=0$ given by Eq.~(\ref{eq:critical_line}).
    }
    \label{fig:critical_line}
\end{figure}

To establish the absolute values of moments $P_k=Q_k$ in the critical system with $σ_0=1$, we further use the obtained equality $δℱ_\aa = δℱ_\bb = 0$. This means that there is a critical line such that the functional $ℱ(\MG + δ\Mg, \MGb + δ\Mgb)$ does not change if (i) $δ\Mg$ is given by Eq.~(\ref{eq:dg}) and $δ\Mgb$ has the corresponding optimal value, or (ii) $δ\Mgb$ is given by Eq.~(\ref{eq:dgb}) and $δ\Mg$ has the corresponding optimal value. Since the critical eigenvalue is nondegenerate and the remaining directions have positive curvature, the local manifold of minima is one-dimensional. Therefore, the two parameterizations obtained by minimizing with respect to $\Mgb$ and $\Mg$, respectively, describe the same critical line. Along this line we can write simultaneously
\begin{spalign}
    δ\Mg = δa\,\FMR\,\MS_{00},
    \\
    δ\Mgb = δb\,\FMRb\,\MSb_{00}
\end{spalign}
if we choose the relation between the coefficients $δa$ and $δb$ appropriately. We call the pair of operators $\MS_{00}$ and $\MSb_{00}$ at $σ_0=1$ the \emph{critical disorder-mediated response channel} or simply the critical channel.

In this case $ℱ(\MG + δ\Mg, \MGb + δ\Mgb) = ℱ^{(0)} + δℱ$ with
\begin{equation}
    δℱ = \frac{(δa + δb)^2}{2} + \sum_{k=3}^{∞}\frac{(-1)^k}{k} \Big(δa^k  + δb^k\Big)P_k.
\end{equation}
For small $δa$, we can neglect higher-order terms and obtain $δb ≈ -δa$. We therefore represent the relation as a power series,
\begin{equation}
    δb = -δa + \sum_{k=2}^∞ h_k δa^k.
\end{equation}
In this case, the expansion of $δℱ$ in powers of $δa$ starts at fourth order
\begin{multline}
    δℱ = \frac{h_2^2 - 2h_2P_3 + P_4}{2}δa^4
    \\
    + \big(h_2 h_3 + (h_2^2 - h_3)P_3 - h_2P_4\big)δa^5 + \ldots
\end{multline}
On the critical line, the coefficients of this expansion must vanish, including the first one. This implies that
\begin{equation}
    h_2 = P_3 \pm \sqrt{P_3^2 - P_4}.
\end{equation}
A real solution exists only if
\begin{equation}
    P_3^2 \ge P_4.
\end{equation}
This inequality is precisely the reverse of (\ref{eq:Pn_ineq}) for $k=3$, given that $P_2 = 1$. Consequently, we obtain the equality
\begin{equation}
     P_3^2 = P_4 P_2.
\end{equation}
Equality in Eq.~(\ref{eq:Pn_ineq}) implies that all nonzero eigenvalues $λ_i$ are equal. Using $P_2=1$, their common value is $1/P_1$, and their number is $P_1^2$. The same reasoning applies to the eigenvalues $μ_i$. Since $P_3=Q_3$, it also follows that $P_1=Q_1$. Consequently, the following matrices satisfy the simple relation that taking the $k$th power is equivalent to multiplication by $P_1^{1-k}$:
\begin{spalign}
    \label{eq:S_pow_crit}
    \big(\MG^{1/2}\thin\MS_{00}\,\MG^{1/2}\big)^k &= P_1^{1-k} \big(\MG^{1/2}\thin\MS_{00}\,\MG^{1/2}\big),
    \\
    \big(\MGb^{1/2}\thin\MSb_{00}\,\MGb^{1/2}\big)^k &= P_1^{1-k}\big(\MGb^{1/2}\thin\MSb_{00}\,\MGb^{1/2}\big)
\end{spalign}
for all integers $k \ge 1$. This observation immediately determines the values of all moments in the critical case $σ_0 = 1$:
\begin{equation}
     P_k = Q_k = P_1^{2-k} = Q_1^{2-k},
\end{equation}
which coincides with the lower bound given in (\ref{eq:PQ_lower_bound}). Such coefficients result in a simple analytical form for the increase in the functional
\begin{multline}
    \label{eq:delta_F_crit}
    δℱ = -P_1^2 + (P_1 + δa)(P_1 + δb)
    \\
    - P_1^2\ln \frac{(P_1 + δa)(P_1 + δb)}{P_1^2}.
\end{multline}
The critical line $δℱ = 0$ is given by the following relation between $δa$ and $δb$
\begin{equation}
    \label{eq:critical_line}
    (P_1 + δa)(P_1 + δb) = P_1^2.
\end{equation}
The increase in the functional, together with the critical line, is shown in Fig.~\ref{fig:critical_line}.

\subsection{Orthogonality preservation of the critical channel}
\label{app:mixing}

Let $\M V$ and $\M V_\bb$ be arbitrary operators that are orthogonal to the critical channel,
\begin{spgather}
    \Tr{\MS_{00}\thin\MG\,\M V\thin\MG} = 0,
    \\
    \Tr{\MSb_{00}\thin\MGb\M V_\bb\thin\MGb} = 0.
\end{spgather}
Consider now the perturbed Green functions along the critical line
\begin{spalign}
    \label{eq:gg_crit}
    \Mg &= \MG + δa\,\FMR\,\MS_{00},
    \\
    \Mgb &= \MGb + δb\,\FMRb\,\MSb_{00}
\end{spalign}
with coefficients $δa > -P_1$ and $δb > -P_1$ that satisfy the relation (\ref{eq:critical_line}). Using Eq.~(\ref{eq:S_pow_crit}), we obtain the orthogonality relation with respect to any $\Mg$ and $\Mgb$ given by Eq.~(\ref{eq:gg_crit})
\begin{spgather}
    \label{eq:gg_crit_orthog}
    \Tr{\MS_{00}\,\Mg\,\M V\,\Mg} = 0,
    \\
    \Tr{\MSb_{00}\,\Mgb\thin\M V_\bb\,\Mgb} = 0.
\end{spgather}
These relations demonstrate that a perturbation taken along the critical line \emph{does not induce mixing} between the critical channel and the other zero-wavevector channels. In particular, one finds
\begin{spgather}
    \Tr{\MS_{00}\thin\MG\,\MS_{00}\,\MG\,\M V\thin\MG} = 0,
    \\
    \Tr{\MSb_{00}\thin\MGb\,\MSb_{00}\,\MGb\,\M V_\bb\thin\MGb} = 0,
\end{spgather}
which means that the following nonlinear coefficients given by (\ref{eq:PQ}) are zero
\begin{equation}
    P^{00n}_{000} = Q^{00n}_{000} = 0
\end{equation}
for $n>0$ when $σ_0 = 1$.

\subsection{Wavevector-dependent channel mixing for a critical system}

We now establish a relation between the nonlinear coefficients containing the zero-wavevector channel ${(n=0,\Vp=0)}$ of the upper branch and two mutually conjugate finite-wavevector channels $(n,\Vp)$ and $(n,-\Vp)$. For the critical system with $σ_0=1$, Eq.~(\ref{eq:S_pow_crit}) gives
\begin{equation}
    \MS_{00}\MG\MS_{00}=\frac{1}{P_1}\MS_{00},
    \qquad
    \MSb_{00}\MGb\MSb_{00}=\frac{1}{P_1}\MSb_{00},
    \label{eq:S_crit_square}
\end{equation}
The critical eigenvalue equations (\ref{eq:S_eigval}) read
\begin{spalign}
    \left\langle
        \MA\MGb\MSb_{00}\MGb\MA^\dag
    \right\rangle_{\!\MA}
    &=\MS_{00},
    \\
    \left\langle
        \MA^\dag\MG\MS_{00}\MG\MA
    \right\rangle_{\!\MA}
    &=\MSb_{00}.
    \label{eq:S_crit_A}
\end{spalign}

Consider now the operator
\begin{equation}
    \MY=\MS_{00}\MG\MA-\MA\MGb\MSb_{00}.
\end{equation}
Since $\MG$ and $\MGb$ are positive definite,
\begin{equation}
    \Tr{\MY^\dag\MG\MY\MGb}\geq0
\end{equation}
for every realization of $\MA$, with equality only if $\MY=0$.
Using Eqs.~(\ref{eq:S_crit_square}) and
(\ref{eq:S_crit_A}), together with $Q_2=1$, we obtain
\begin{multline}
    \left\langle
        \Tr{\MY^\dag\MG\MY\MGb}
    \right\rangle_{\MA}
    =
    \frac{1}{P_1}\Tr{\MSb_{00}\MGb}
    \\
    -2\Tr{\MSb_{00}\MGb\MSb_{00}\MGb}
    +\frac{1}{P_1}\Tr{\MS_{00}\MG} = 0.
\end{multline}
Since the quantity inside the average is nonnegative, it must vanish
for almost every realization of $\MA$. Therefore,
\begin{equation}
    \label{eq:A_intertwining}
    \MS_{00}\MG\MA
    =
    \MA\MGb\MSb_{00}.
\end{equation}
Taking the Hermitian conjugate gives the equivalent relation
\begin{equation}
    \MA^\dag\MG\MS_{00}
    =
    \MSb_{00}\MGb\MA^\dag.
\end{equation}

We can now use Eq.~(\ref{eq:A_intertwining}) to relate the nonlinear
coefficients in the two operator spaces. For an arbitrary
disorder-mediated channel $(n,\Vp)$, the first
eigenvalue equation~(\ref{eq:S_eigval}) gives
\begin{align}
    σ_{n\Vp}P^{0\,n\,n}_{0,\Vp,-\Vp}
    &=
    \left\langle
        \Tr{
            \MG\MS_{00}\MG\MA
            \MGb\MSb_{n\Vp}\MGb\MA^\dag
            \MG\MS_{n,-\Vp}
        }
    \right\rangle_{\!\MA}
    \notag\\
    &=
    \left\langle
        \Tr{
            \MG\MA\MGb\MSb_{00}\MGb\MSb_{n\Vp}
            \MGb\MA^\dag\MG\MS_{n,-\Vp}
        }
    \right\rangle_{\!\MA},
\end{align}
where Eq.~(\ref{eq:A_intertwining}) was used in the second line.
Using cyclic invariance of the trace, this expression can be written as
\begin{multline}
    σ_{n\Vp}P^{0\,n\,n}_{0,\Vp,-\Vp}
    =
    \Trr{
        \MGb\MSb_{00}\MGb\MSb_{n\Vp}\MGb
        \left\langle
            \MA^\dag\MG\MS_{n,-\Vp}\MG\MA
        \right\rangle_{\!\MA}
    }.
\end{multline}
The second eigenvalue equation~(\ref{eq:S_eigval}), together with
$σ_{n,-\Vp}=σ_{n\Vp}$, gives
\begin{equation}
    \left\langle
        \MA^\dag\MG\MS_{n,-\Vp}\MG\MA
    \right\rangle_{\!\MA}
    =
    σ_{n\Vp}\MSb_{n,-\Vp}.
\end{equation}
Consequently,
\begin{equation}
    σ_{n\Vp}P^{0\,n\,n}_{0,\Vp,-\Vp}
    =
    σ_{n\Vp}Q^{0\,n\,n}_{0,\Vp,-\Vp}.
\end{equation}
Since $σ_{n\Vp}>0$, we finally obtain the equality of the wavevector-dependent
channel-mixing coefficients,
\begin{equation}
    \label{eq:Ppp_Qpp}
    P^{0\,n\,n}_{0,\Vp,-\Vp}
    =
    Q^{0\,n\,n}_{0,\Vp,-\Vp}
\end{equation}
in the critical system with $σ_0=1$.

\section{Nonlinear kernel}

\label{app:nl_kernel}

We derive the long-wavelength expansion of the upper-branch kernels
$P^{0\,0\,0}_{\Vp\,\Vp'\Vp''}$ and
$Q^{0\,0\,0}_{\Vp\,\Vp'\Vp''}$ entering
Eq.~(\ref{eq:nl_kernel_exp}). Owing to translational invariance, we
choose the first spatial argument as the origin $\Vr=0$. With the Fourier convention
\begin{equation}
    \MS_{0\Vr}
    =
    \frac{1}{\sqrt{N}}
    \sum_{\Vp}e^{-i\Vp\Vr}\MS_{0\Vp},
\end{equation}
we then obtain
\begin{spalign}
    P^{0\,0\,0}_{\Vp\,\Vp'\Vp''}
    &=
    \frac{δ_{\Vp+\Vp'+\Vp'',0}}{\sqrt{N}}
    \sum_{\Vr'\Vr''}
    P_{0,\Vr',\Vr''}
    e^{i(\Vp'\Vr'+\Vp''\Vr'')},
    \label{eq:Fourier-normalization}
    \\
    Q^{0\,0\,0}_{\Vp\,\Vp'\Vp''}
    &=
    \frac{δ_{\Vp+\Vp'+\Vp'',0}}{\sqrt{N}}
    \sum_{\Vr'\Vr''}
    Q_{0,\Vr',\Vr''}
    e^{i(\Vp'\Vr'+\Vp''\Vr'')}.
\end{spalign}

For an isotropic and nonchiral ensemble, the first moments
vanish. Translational invariance makes the following moments independent
of $\Vr$, while isotropy allows us to write
\begin{spalign}
    \sum_{\Vr'\Vr''}P_{0,\Vr',\Vr''}&=\frac{1}{a},
    \\
    \sum_{\Vr'\Vr''}Q_{0,\Vr',\Vr''}&=\frac{1}{b}
\end{spalign}
and
\begin{spalign}
    \sum_{\Vr'\Vr''}P_{0,\Vr',\Vr''}r'_\alpha r''_\beta
    &=\frac{\nu_\aa}{a}\delta_{\alpha\beta},
    \\
    \sum_{\Vr'\Vr''}Q_{0,\Vr',\Vr''}r'_\alpha r''_\beta
    &=\frac{\nu_\bb}{b}\delta_{\alpha\beta}.
\end{spalign}
Here $a$, $b$, $\nu_\aa$, and $\nu_\bb$ are four constants.

Cyclic symmetry of the three-point kernels implies that the second
moments of the three relative vectors $\Vr'$, $\Vr''$, and
$\Vr''-\Vr'$ are identical. Together with the mixed moments above,
this gives
\begin{spalign}
    \sum_{\Vr'\Vr''}P_{0,\Vr',\Vr''}r'_\alpha r'_\beta
    &=
    \sum_{\Vr'\Vr''}P_{0,\Vr',\Vr''}r''_\alpha r''_\beta
    =\frac{2\nu_\aa}{a}\delta_{\alpha\beta},
    \\
    \sum_{\Vr'\Vr''}Q_{0,\Vr',\Vr''}r'_\alpha r'_\beta
    &=
    \sum_{\Vr'\Vr''}Q_{0,\Vr',\Vr''}r''_\alpha r''_\beta
    =\frac{2\nu_\bb}{b}\delta_{\alpha\beta}.
\end{spalign}

Expanding Eqs.~(\ref{eq:Fourier-normalization}) to second order and using
$\Vp+\Vp'+\Vp''=0$, we obtain
\begin{spalign}
    P^{0\,0\,0}_{\Vp\,\Vp'\Vp''}
    &=\frac{δ_{\Vp+\Vp'+\Vp'',0}}{a\sqrt{N}}
      \left[1-\frac{ν_\aa}{2}
      \bigl(|\Vp|^2+|\Vp'|^2+|\Vp''|^2\bigr)\right],
    \\
    Q^{0\,0\,0}_{\Vp\,\Vp'\Vp''}
    &=\frac{δ_{\Vp+\Vp'+\Vp'',0}}{b\sqrt{N}}
      \left[1-\frac{ν_\bb}{2}
      \bigl(|\Vp|^2+|\Vp'|^2+|\Vp''|^2\bigr)\right].
    \label{eq:PQ-small-wavevector}
\end{spalign}
Equation~(\ref{eq:Ppp_Qpp}) with $n=0$ implies $a=b$ and
$ν_\aa=ν_\bb$ at $σ_0=1$. Therefore, the corresponding differences are small in the near-critical regime $1-σ_0 \ll 1$.

For fields that vary slowly over the range of the kernels, we restore
the arbitrary center $\Vr$ and perform a Taylor expansion, obtaining
\begin{align}
    &\sum_{\Vr'\Vr''}P_{\Vr,\Vr',\Vr''}
    f(\Vr')g(\Vr'')
    =
    \frac{1}{a}\bigl[fg+\nu_\aa\mathcal{D}(f,g)\bigr]_{\Vr},
    \\
    &\sum_{\Vr'\Vr''}Q_{\Vr,\Vr',\Vr''}
    f(\Vr')g(\Vr'')
    =
    \frac{1}{b}\bigl[fg+\nu_\bb\mathcal{D}(f,g)\bigr]_{\Vr}
\end{align}
where $\mathcal{D}(f,g)=(∇f)\cdot(∇g)+fΔg+gΔf$. This reproduces
Eq.~(\ref{eq:nl_kernel_exp}).

\section{Calculation of the nonlocal elastic kernels}
\label{app:nonlocal}

To determine the nonlocal elastic response, we study the atomic displacements $\V{u}_i$ induced by weak external forces $\V{f}_i$. At the continuum level, these forces define a body-force density $\V{b}(x)$, while the atomic displacements define a displacement field $\V{u}(x)$. Since the perturbations considered in the main text depend only on the coordinate $x$ normal to the confining planes, mechanical equilibrium is expressed as
\begin{equation}
    -\frac{∂σ_{αx}(x)}{∂x}=b_α(x).
    \label{eq:nonlocal-equilibrium}
\end{equation}
Under the assumption of an isotropic two-modulus representation of
the elastic response, the most general nonlocal constitutive relation
relevant to this geometry can be written as
\begin{multline}
    σ_{αx}(x)
    =
    \int 
    \biggl[ δ_{αx}δ_{βx}
    \left( \KK(x,x')+\frac{1}{3}\GG(x,x') \right) 
    \\
    + δ_{αβ}\GG(x,x') \biggr]
    \frac{∂u_β(x')}{∂x'}\,\dd x'.
    \label{eq:nonlocal-constitutive-general}
\end{multline}
Here, $\KK(x,x')$ and $\GG(x,x')$ are the nonlocal generalizations of the bulk and shear moduli, respectively~\cite{Eringen-nonlocal-continuum-field-2004}. Equation~\eqref{eq:nonlocal-constitutive-general} shows that transverse deformations are governed by $\GG(x,x')$, whereas longitudinal deformations are governed by the longitudinal modulus kernel
\begin{equation}
    \mathsf{M}(x,x')
    =
    \KK(x,x')+\frac{4}{3}\GG(x,x').
\end{equation}
To determine these kernels for the molecular-dynamics configurations, we apply spatially modulated force fields, average the resulting atomic displacements over the available configurations, and reconstruct the corresponding longitudinal and transverse response operators as described below.

We consider an amorphous slab occupying the interval $x\in[-W/2,W/2]$. The slab is confined between two parallel rigid planes at $x=-W/2$ and $x=W/2$, whose area is denoted by $A_x$. Throughout the calculation, the planes remain fixed. The reference configuration is a mechanically stable local minimum of the potential energy at $T=0$.

It is convenient to introduce the wave numbers $q_n = n π/W$ for $n=1,2,\ldots$ and the corresponding basis functions
\begin{align}
    s_n(x) &= \sin [q_n(x+W/2)],
    \\
    c_n(x) &= \cos [q_n(x+W/2)].
\end{align}
The functions $s_n(x)$ vanish at both rigid planes and therefore satisfy the boundary conditions for the displacement field.

For a polarization direction $α\in\{x,y,z\}$ and a mode number $n$, we apply to each mobile atom $i$ the external force
\begin{equation}
    \V{f}_i^{(α,n)}
    = \V{e}_α\,\frac{σ_{\mathrm m}A_x}{N_{\mathrm{at}}}\,s_n(x_i),
    \label{eq:nonlocal-atomic-force}
\end{equation}
where $N_{\mathrm{at}}$ is the number of mobile atoms and $σ_{\mathrm m}$
sets the stress scale. If the mean atomic number density is
$N_{\mathrm{at}}/(A_xW)$, Eq.~\eqref{eq:nonlocal-atomic-force} corresponds to
the continuum body-force density
\begin{equation}
    \V{b}^{(α,n)}(x)
    = \V{e}_α\,\frac{σ_{\mathrm m}}{W}\,s_n(x).
    \label{eq:nonlocal-body-force}
\end{equation}
Thus, the amplitude of the $n$th body-force mode is
$b_n=σ_{\mathrm m}/W$.

For each applied force field, the energy is minimized using the FIRE algorithm
while the confining planes are held fixed. In the linear-response regime, the resulting
displacement of atom $i$ is denoted by $\V{u}_i^{(α,n)}$. Its projection onto
the sine basis is
\begin{equation}
    U_{kβ}^{(α,n)}
    = \frac{2}{N_{\mathrm{at}}}
      \sum_{i=1}^{N_{\mathrm{at}}}
      s_k(x_i)\,u_{iβ}^{(α,n)},
    \qquad α\in\{x,y,z\}.
    \label{eq:nonlocal-displacement-projection}
\end{equation}
For a statistically uniform distribution of atoms along $x$, this discrete projection is the particle representation of $(2/W)\int_{-W/2}^{W/2}s_k(x)u_β(x)\,\dd x$. 

We define the longitudinal and transverse compliance matrices by
\begin{align}
    \widetilde{L}_{\mathrm L,kn}
    &= \frac{W}{σ_{\mathrm m}}\,U_{kx}^{(x,n)},
    \label{eq:nonlocal-longitudinal-compliance}\\
    \widetilde{L}_{\mathrm T,kn}
    &= \frac{W}{2σ_{\mathrm m}}
       \left(U_{ky}^{(y,n)}+U_{kz}^{(z,n)}\right).
    \label{eq:nonlocal-transverse-compliance}
\end{align}
The first index labels the displacement mode and the second index labels the
applied-force mode. These matrices relate the modal displacement and
body-force amplitudes according to
\begin{equation}
    u_k = \sum_n \widetilde{L}_{kn}b_n.
    \label{eq:nonlocal-modal-compliance}
\end{equation}

Within the isotropic two-modulus representation used here, the transverse
response determines the shear kernel $\GG(x, x')$, whereas the longitudinal response
determines $\mathsf{M}(x,x')$. The corresponding matrices in the cosine basis are obtained by inverting the
compliance matrices:
\begin{align}
    \widetilde{\GG}_{nm}
    &= \frac{1}{q_nq_m}
       \left(\widetilde{L}_{\mathrm T}^{-1}\right)_{nm},
    \label{eq:nonlocal-G-modal}\\
    \widetilde{\mathsf{M}}_{nm}
    &= \frac{1}{q_nq_m}
       \left(\widetilde{L}_{\mathrm L}^{-1}\right)_{nm},
    \label{eq:nonlocal-M-modal}\\
    \widetilde{\KK}_{nm}
    &= \widetilde{\mathsf{M}}_{nm}
       -\frac{4}{3}\widetilde{\GG}_{nm}.
    \label{eq:nonlocal-K-modal}
\end{align}

The real-space kernels are reconstructed using the inverse cosine transform,
\begin{align}
    \GG(x,x')
    &= \frac{2}{W}\sum_{n,m}
       c_n(x)\,\widetilde{\GG}_{nm}\,c_m(x'),
    \label{eq:nonlocal-G-realspace}\\
    \KK(x,x')
    &= \frac{2}{W}\sum_{n,m}
       c_n(x)\,\widetilde{\KK}_{nm}\,c_m(x').
    \label{eq:nonlocal-K-realspace}
\end{align}
In numerical calculations, both sums are truncated at the largest mode for
which the response can be determined reliably.

The applied sine modes generate displacement fields that vanish at both
planes. Consequently, their strain fields have zero spatial average, and the
procedure probes only the subspace spanned by $c_n(x)$ with $n\geq1$. The
spatially uniform strain mode is therefore not determined by this method. We determine the unknown coefficients $\widetilde{\GG}_{0n}$ and $\widetilde{\GG}_{n0}$ by requiring that $\GG(x, x')$ decays to zero for large separations $|x-x'|$. The same procedure is applied to calculate $\KK(x, x')$.

\bibliography{refs1.bib}

\end{document}